\documentclass[twocolumn,superscriptaddress,longbibliography]{revtex4-2}
\usepackage[T1]{fontenc}
\usepackage{times}
\usepackage[utf8]{inputenc}
\usepackage[resetlabels,labeled]{multibib}
\usepackage{enumitem}
\usepackage{amsmath}
\usepackage{amsthm}
\usepackage{amsfonts}
\usepackage{amssymb}
\usepackage{bm}
\usepackage{bbm}
\usepackage{bbold}
\usepackage{tikz}
\usepackage{pgfplots}
\usepackage{graphicx,epsfig,color}
\usepackage{physics}
\usepackage{floatrow}
\usepackage{float} 
\usepackage{mathtools}
\usepackage{mathrsfs}
\usepackage{latexsym}
\usepackage[colorlinks=true,linkcolor=teal,citec olor=purple]{hyperref}
\usepackage{cleveref}
\usepackage[mathscr]{eucal}
\usepackage{extarrows}
\usepackage{tikz}
\usepackage[ruled,vlined]{algorithm2e}
\usepackage{bm}

\newcommand{\chg}[1]{\color{black}#1 \color{black}}
\newcommand{\ot}{\otimes}
\newcommand{\ms}[1]{\mathcal{#1}}

\SetKwInput{KwInput}{Input}               
\SetKwInput{KwOutput}{Output}             
\SetKwInput{KwParams}{Parameters}

\pgfplotsset{compat=1.18} 

\begin{document}

\title{Thermodynamic Computing via Autonomous Quantum Thermal Machines}
\author{Patryk Lipka-Bartosik} 
 \affiliation{Department of Applied Physics, University of Geneva, 1211 Geneva, Switzerland}
\author{Mart\'i Perarnau-Llobet}
 \affiliation{Department of Applied Physics, University of Geneva, 1211 Geneva, Switzerland}
 \author{Nicolas Brunner}
  \affiliation{Department of Applied Physics, University of Geneva, 1211 Geneva, Switzerland}

\begin{abstract}
We develop a physics-based model for classical computation based on autonomous quantum thermal machines. These machines consist of few interacting quantum bits (qubits) connected to several environments at different temperatures. Heat flows through the machine are here exploited  for computing.  The  process starts by setting the temperatures of the environments according to the logical input. The machine evolves, eventually reaching a non-equilibrium steady state, from which the output of the computation can be determined via the temperature of an auxilliary finite-size reservoir. Such a machine, which we term a ``thermodynamic neuron'', can implement any linearly-separable function, and we discuss explicitly the cases of NOT, 3-MAJORITY and NOR gates. In turn, we show that a network of thermodynamic neurons can perform any desired function. We discuss the close connection between our model and artificial neurons (perceptrons), and argue that our model provides an alternative physics-based analogue implementation of neural networks, and more generally a platform for thermodynamic computing. 
\end{abstract}

\maketitle

\section{Introduction}

Computing systems can take a variety of forms, from biological cells to massive supercomputers, and perform a broad range of tasks, from basic logic operation to machine learning. In all cases, the computational process must adhere to the principles of physics, and, in particular, to the laws of thermodynamics. In general information processing and thermodynamics are deeply connected, see e.g. \cite{bennett1982thermodynamics,parrondo2015thermodynamics,wolpert2019stochastic}. 

More recently, links between thermodynamics and computation are being developed. At the fundamental level, bounds for the thermodynamic cost of computation have been derived, see e.g. \cite{Deffner2013,Boyd2018,Faist2018}. From a more practical perspective, a promising direction explores low-dissipation computing. Here, models for elementary gates and circuits based on electronic transistors working in the mesoscopic regime, or even towards the single-electron mode, are considered \cite{gu2019microreversibility, Gu2019,wolpert2020thermodynamics,Gao2021,Freitas_2021,helms2022stochastic,Kuang2022Modelling}. \chg{Crucially, thermodynamic models of computation must be thermodynamically consistent, meaning they adhere to the laws of thermodynamics~\cite{seifert2012stochastic}. This allows to analyze their thermodynamic properties, e.g. energetic cost or dissipated heat, using the framework of stochastic thermodynamics.} This approach already brought considerable progress and further insight can be expected by moving to the fully quantum regime~\cite{Solfanelli_2022,PRXQuantum.2.040335,fellousasiani2022optimizing,Auff_ves_2022,Stevens_2022,Solfanelli_2022}.

Another exciting direction is thermodynamic computing~\cite{Conte2023,coles2023thermodynamic,aifer2024error,duffield2023thermodynamic}. This represents a paradigm  for alternative physics-based models of computation, similarly to quantum computing or DNA computing. The main idea is to exploit the thermodynamic behaviour of complex, non-equilibrium physical systems to perform  computations, looking for a computational speed-up but also a reduced energy cost. This approach has been explored in the context of machine learning and AI, see e.g.~\cite{Goldt2017,e22030256,Hylton2022,Boyd2022}. Very recently, promising progress has been reported, showing that a computational speedup in  linear algebra problems can be achieved via a controllable system of coupled harmonic oscillators embedded in a thermal bath~\cite{aifer2023thermodynamic}.

\begin{figure}
    \centering
    \includegraphics[width=\linewidth]{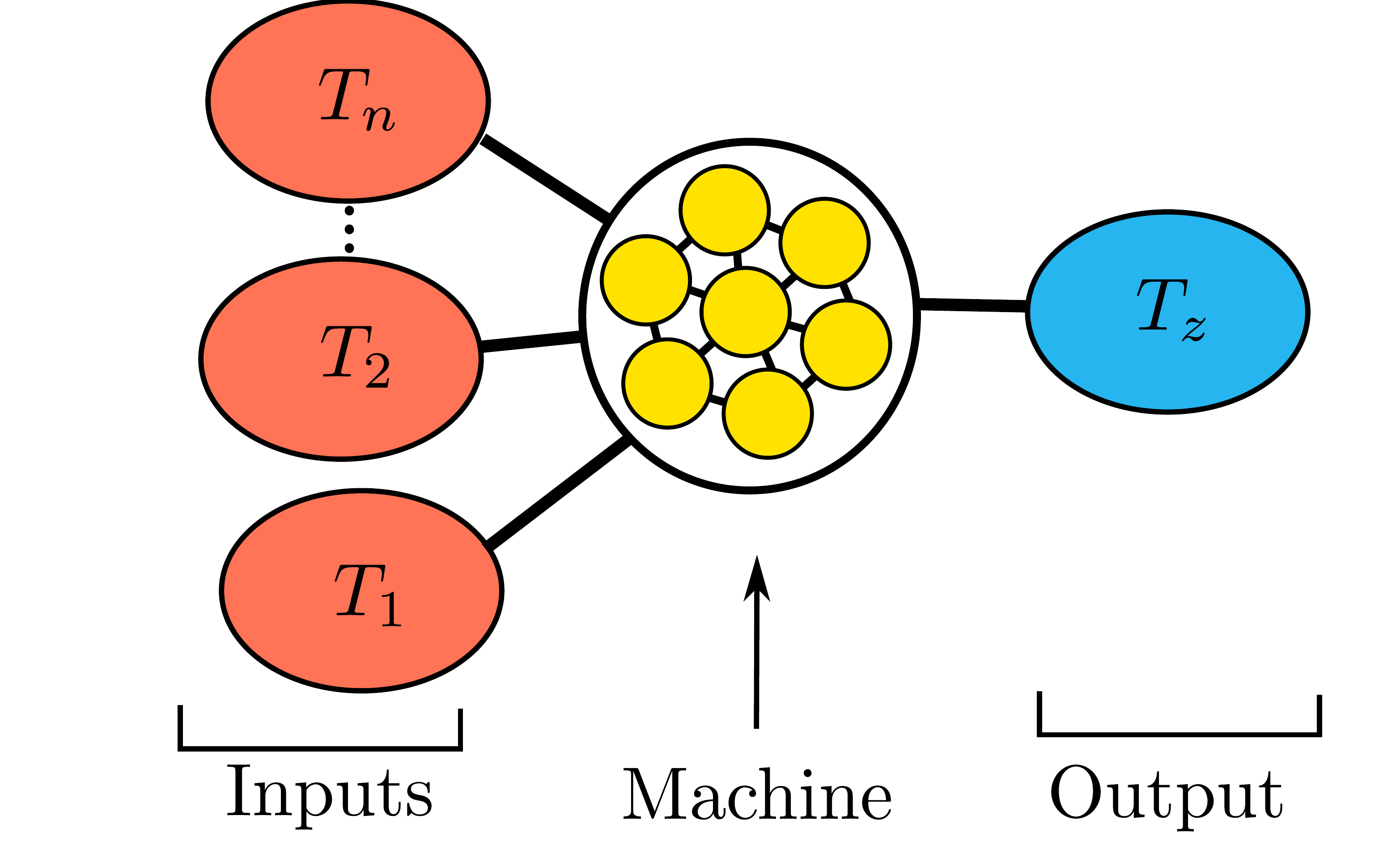}
    \caption{\textbf{Thermodynamic neuron.} The thermodynamic neuron is an autonomous quantum thermal machine designed for computing. The device consists of few interacting qubits (yellow dots), connected to several thermal environments. The input of the computation is encoded in the temperature of heat baths (depicted in red). This generates heat flows through the machine, which eventually reaches a non-equilibrium steady state. The output of the computation can be retrieved from the final temperature of a finite-size reservoir (shown in blue). By designing the machine (setting the qubit energies and their interaction), specific functions between the input and output temperatures can be implemented.   
    }
    \label{fig:scheme}
\end{figure}

In this work, we develop a model for thermodynamic computing starting from a minimal model of a quantum thermal machine.  
More precisely, we develop autonomous quantum thermal machines  that  can operate as computing devices where logical inputs and outputs are encoded in the temperature.  As our device shares strong similarities with the basic model of an artificial neuron (the perceptron used e.g. in neural networks), we refer to it as a ``thermodynamic neuron''. Overall, our guiding motivation is to use diverse techniques offered by quantum thermodynamics to enhance our understanding of  fundamental aspects of computation.

To construct our computing device, we start from the model of minimal autonomous quantum thermal machines~\cite{Linden_2010,Brunner2012}, which are made of a small quantum system (few interacting qubits) in contact with thermal baths at different temperatures. A first observation is that the effect of such a thermal machine onto an external system---heating or cooling---depends on the temperatures of the heat baths. Viewing these temperatures as an input, and the temperature of the external system as an output, the thermal machine can be seen as a computing device (see Fig. \ref{fig:scheme}). By associating a logical value to the temperature (e.g. cold temperatures corresponding to logical ``0'' and hot temperatures to logical ``1''), we show that the autonomous machine can implement logical gates. As a first example, we show how a small quantum refrigerator/heat pump can be used to implement an inverter (NOT gate). This represents the simplest example of a thermodynamic neuron. In turn, we present a general model of a thermodynamic neuron, and show that it can implement any \chg{Boolean} linearly-separable function. \chg{Such a function can be thought of as an assignment of $0$ or $1$ to the vertices of a Boolean hypercube (i.e. a geometric representation of its truth table). This allows to divide the vertices into two sets. The Boolean function is said to be linearly separable if these two sets of points can be separated with a line.}. We discuss explicitly the examples of NOR and 3-MAJORITY. A key element in this construction are the concepts of virtual qubits and virtual temperatures \cite{Brunner2012}, which allow us to establish a close connection between our machines and perceptrons, a common model of an artificial neuron. Furthermore, we show that by constructing networks of thermodynamic neurons, one can implement any desired function, and we discuss the example of XOR. We detail an algorithm, inspired by artificial neural networks, for designing thermodynamic neurons (and networks of them) for implementing any given target function. We conclude with a discussion and an outlook.

Before proceeding, we highlight a number of relevant features of our model. First, as it is constructed from a minimal model of quantum thermal machines, the model is thermodynamically consistent. Hence the model allows for \chg{an examination of} the trade-off between consumed energy, dissipation and performance, which we investigate. Second, as it is based on changes of temperatures and flows of energy, the model involves only one conserved quantity, namely energy. \chg{Computation in our model occurs solely as a result of heat flowing from one part of the machine to the other. This is in contrast to most conventional models of computation, in particular models for nano-scale electronic circuits, where heat is an unwanted byproduct that hampers computation and introduces errors.} Finally, the functioning of our model can be intuitively understood by exploiting interesting connections between quantum systems at thermal equilibrium and artificial neural networks. 

\section{Results}
\subsection{Framework}
\subsubsection{Autonomous quantum thermal machines} \label{sec:methods}

Quantum thermal machines usually consists of a small-scale physical system described within quantum theory. This system is then placed in contact with external resources, such as thermal baths or driving, in order to implement a thermodynamic task such as cooling, heating or producing work; see e.g.~\cite{Goold2015} or~\cite{Vinjanampathy_2016} for reviews.

Here our focus is on a special class of quantum thermal machines known as autonomous quantum thermal machines (see~\cite{Mitchison2019} for recent review). Their main interest resides in the fact that these machines work autonomously, in the sense that they are powered by external resources that thermal (typically two or more heat baths at different temperatures) and their internal dynamics is time-independent (modeled via a time-independent Hamiltonian). While first models can be traced back to the thermodynamic analysis of masers~\cite{SD1959}, recent works have developed a framework for discussing minimal models of autonomous thermal machines, working as refrigerators, heat pumps and heat engines~\cite{Linden_2010,Brunner2012,levy2012quantum}. Interestingly many physical models of quantum thermal machines~\cite{PhysRevLett.110.256801,correa2014quantum,hofer2016,PhysRevE.87.012140,PhysRevLett.126.180605,Niedenzu2019conceptsofworkin} can be mapped back to these minimal abstract models~\cite{Mitchison2019}. More recently, autonomous machines have also been devised for achieving other tasks such as the creation of entanglement~\cite{Bohr_Brask_2015}, time-keeping (i.e. clocks) \cite{Erker2017,PhysRevX.11.011046,Woods2021} and thermometry \cite{Hofer2017}. A key aspect of these machines is their autonomy making them relevant from a practical perspective~\cite{guzmán2023divincenzolike}, and first proof-of-principle experiments have been reported~\cite{Maslennikov2019,aamir2023thermally}. More generally, the limits of designing autonomous quantum devices have been discussed~\cite{Woods2023}.

\subsubsection{Open quantum system dynamics}

In this work, we will focus on autonomous quantum thermal machines consisting of few \emph{qubits}, i.e few two-level quantum systems. To start with, let us review the dynamics of a single qubit in contact with a heat bath. First, the qubit features two energy eigenstates: the ground state $\ket{0}$ and the excited state $\ket{1}$, with respective energies $E_0$ and $E_1>E_0$. The state of the qubit is represented by a density operator $\rho$, and its mean energy is given by $\Tr[\rho H]$, where $H = E_0 \dyad{0} + E_1 \dyad{1}$ denotes the Hamiltonian. A convenient quantity is the energy gap, $\epsilon := E_1 - E_0$. Without loss of generality we take $E_0 = 0$, so that the qubit's energy is fully specified by its energy gap. When placed in contact with an environment, the qubit evolution is described by the \emph{master equation}:
\begin{align}
    \dot{\rho} = - i [H, \rho] + \mathcal{L}[\rho].
\end{align}
The first term captures the unitary evolution governed by the Hamiltonian, while the second term captures the environment's impact on the qubit via the dissipator $\mathcal{L}[\cdot]$. Here we use the common assumption of weak coupling to write down the dissipator, i.e. we assume that the qubit is weakly correlated with its environment.

As the qubit evolves over time, it eventually reaches a steady-state when $\dot{\rho} = 0$. When the environment is a thermal bath, with an inverse temperature $\beta = 1/kT$, the resulting steady-state is given by a qubit thermal (Gibbs) state: $\tau(\beta) = e^{-\beta H} / Z$, where $Z = \tr e^{-\beta H}$ is the canonical partition function. In this case, the probability of the qubit to be in the excited state is given by the Fermi-Dirac distribution
\begin{align}\label{eq:fd}
    g(\beta \epsilon) = \braket{1}{\tau(\beta)|1} = \frac{1}{1 + e^{\beta \epsilon}}.
\end{align}
Note that this function coincides with the sigmoid function used in machine learning. We will explore this connection more carefully later.

\subsubsection{Thermal machines}

The machines we will consider typically consist of several qubits with energy gaps $\epsilon_k$. The qubits weakly interact with each other, via an energy-preserving interaction. This is modeled by a time-independent interaction Hamiltonion, $H_{\text{int}}$, which commutes with the free Hamiltonian $H_0 = \sum_k \epsilon_k \dyad{1}_k$, i.e. $[H_{\text{int}},H_0]=0$. \chg{In what follows we will slightly abuse notation and write $\dyad{i}_k$ to denote a tensor product acting as identity everywhere except at position $k$, i.e. $\mathbb{1} \otimes \ldots \otimes \dyad{i}_k  \otimes \ldots \otimes \mathbb{1}$.} Each qubit is then connected to a thermal bath. In general these baths are at different (inverse) temperatures $\beta_k$. \chg{When the coupling between  qubits and thermal baths is weak}, the dynamics of such a machine is well captured by a local master equation \cite{hofer2017markovian} of the form
\begin{align}
    \label{eq:master}
    \dot{\rho} = - i [H_0 + H_{\text{int}}, \rho] + \sum_k \mathcal{L}^{(k)}[\rho_k].
\end{align}
where $\rho$ now denotes the multi-qubit state of the machine.

\chg{The main assumption which we are going to use in this work is  local \emph{detailed balance}  
which, in our current context, means that \emph{local} thermal states are the fixed point of each dissipator, i.e.
\begin{align}
    \mathcal{L}^{(k)}[\tau(\beta_k)] = 0.
\end{align}
This condition is well justified when the couplings in $H_{\rm int}$ are sufficiently weak~\cite{hofer2017markovian}.
A quantity relevant to our analysis is the heat current \chg{released from the qubit to the heat bath in this process. This is given by 
\begin{align}
j_{k} := \text{Tr}[H \mathcal{L}^{(k)}[\rho]]. 
\end{align}}
We note that in certain cases, a qubit of the machine will be coupled to two different baths, in general at different temperatures. In this case, the total dissipators for the qubit is simply obtained by summing the dissipators with respect to each bath. In turn, this implies that the total heat current is the sum of the heat currents with respect to each bath.

}
\chg{Although our key qualitative findings only require the detailed balance condition, introducing a specific thermalization model would allow us to support our results with numerical evidence. To keep the presentation simple} we will use the so-called \emph{reset model} (see e.g. \cite{Linden_2010}) in which the dissipators take the simple form
   $\mathcal{L}^{(k)}[\rho] = \gamma_{k} \left(\Tr_k[\rho] \ot \tau(\beta_k) -\rho \right)$,
\chg{where $\Tr_k[\cdot]$ denotes the partial trace over qubit $k$} and $\gamma_{k}$ is the coupling, which corresponds to the probability that qubit $k$ thermalizes with its bath. \chg{We assume that all systems are labelled, and no relevance is given to the order of the tensor product.} Note that $\Tr_k[\rho] \ot \tau_{\chg{k}}(\beta_k)$ represents the multi-qubit state after \chg{a full} thermalization event. This model can be viewed as a collisional process, where in each instant of time the qubit has a certain probability to collision with a thermal qubit from the bath. Within this model the heat current from Eq. \eqref{eq:reset} is given by 
\begin{align}
\label{eq:reset}
j_{k} := \text{Tr}[H \mathcal{L}^{(k)}[\rho]] = \gamma_{k} \epsilon_k \left[g(\beta_k \epsilon_k
) - p_k\right],
\end{align}
where $p_k$ is the probability that the qubit connected to the bath is in an excited state. We note that in certain cases, a qubit of the machine will be coupled to two different baths, in general at different temperatures. In this case, the total dissipators for the qubit is simply obtained by summing the dissipators with respect to each bath. In turn, this implies that the total heat current is the sum of the heat currents with respect to each bath.

Finally, a quantity of interest for our work is the dissipation generated by the machines. To quantify dissipation, we use \chg{entropy production rate} $\dot{\Sigma}$. This quantity captures the fundamental irreversibility of the machine. The second law of thermodynamics restricts the behavior of any thermal machine. For our autonomous machines it reads 
\begin{align}\label{eq:not_gate_1law}
    \dot{\Sigma} := \dot{S}(\rho(t)) - \sum_{k} \beta_k j_k(t) \geq 0,
\end{align}
where $S(\rho) := - \Tr[\rho \log \rho]$ is the von Neumann entropy of the machine and $j_{k}(t)$ is the total heat current flowing into the $k$-th heat bath at time $t$. \chg{We also use the dot notation to indicate complete time derrivatives, e.g. $\dot{\Sigma} \equiv \frac{\text{d}}{\text{d}t} \Sigma$.}

The quantity $\dot{\Sigma}$ is the rate of \emph{entropy production} which quantifies the speed at which heat (entropy) is dumped into all environments connected with the machine, see e.g. \cite{Tolman1948,Goold2015,Landi_2021}. It therefore measures the amount of information that is lost (i.e. transferred to unobserved degrees of freedom). It is also a central quantity appearing in thermodynamic uncertainty relations (TURs) \cite{Horowitz_2017,barato2015thermodynamic,falasco2020unifying}, as well as bounds on the speed of a stochastic evolution \cite{Shiraishi_2018}. \chg{We will be mostly interested in the dynamics of the \emph{steady state} of the system which corresponds to $\dot{\rho} = 0$, or equivalently $\dot{S}(\rho(t)) = 0$.}

\chg{An important class of quantum thermal machines are \emph{autonomous} machines. Such machines operate without requiring external control over their internal components (e.g. couplings or local energies), as they operate in the steady-state regime. This autonomy offers a key advantage: it eliminates the need for complex, high-precision control, which is a major contributor to the energy consumption of traditional nanoscale devices. An interesting platform for realizing autonomous quantum thermal machines are thermoelectric quantum dots \cite{Sothmann_2015}.}

\begin{figure}
    \centering
    \includegraphics[width=\linewidth]{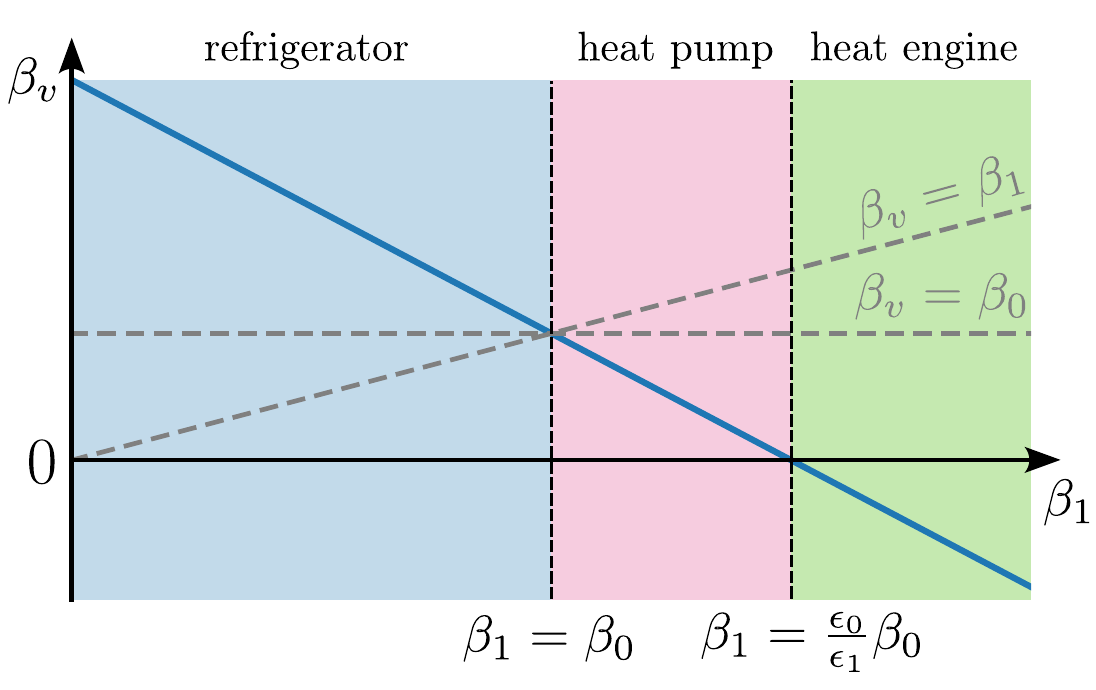}
    \caption{\chg{\textbf{Different operation regimes of a three-qubit thermal machine.} The plot shows the inverse virtual temperature $\beta_v$ as a function of bath inverse temperature $\beta_1$ when keeping $\beta_0$ fixed. When $\beta_1 < \beta_0$, the inverse virtual temperature becomes larger than both $\beta_0$ and $\beta_1$, which means that the machine operates as a refrigerator. When $\beta_0 < \beta_1 < (\epsilon_0/\epsilon_1)\beta_0$ we have the exactly opposite situation and the machine operates as a heat pump. Finally, when $\beta_1 > (\epsilon_0/\epsilon_1)\beta_0$ the machine operates as a heat engine. Figure adapted from Ref. \cite{Brunner2012}.} }
    \label{fig:regimes}
\end{figure}

\chg{
\subsubsection{Autonomous quantum thermal machines and virtual qubit}
\label{sec:three_qubits}
Before we explain our model of a computing thermal machine it is worth discussing a simpler machine, namely the three-qubit thermal machine introduced in Ref. \cite{Linden_2010,Brunner2012}. The intuition developed for this model can be then used to understand more complex quantum thermal machines.

Consider a thermal machine that consists of two qubits $\ms{C}_0$ and $\ms{C}_1$ such that $\ms{C}_i$ is in a thermal contact with a heat bath at an inverse temperature $\beta_i$ for $i = 0, 1$. Let $\epsilon_0$ be the energy spacing of qubit $\ms{C}_0$ and $\epsilon_1 \leq \epsilon_0$ be the energy spacing of qubit $\ms{C}_1$. In the absence of interactions with an external system, each qubit interacts only with its own thermal bath, and hence reaches thermal equilibrium at the corresponding inverse temperature. Therefore the state of qubit $\ms{C}_i$ can be written as
\begin{align}
    \tau_{\ms{C}_i}(\beta_i) = \frac{1}{Z_{\ms{C}_i}} \left(\dyad{0}_{\ms{C}_i} + e^{-\beta_i \epsilon_i }\dyad{1}_{\ms{C}_i} \right),
\end{align}
where $Z_{\ms{C}_i} = 1 + e^{-\beta_i \epsilon_i}$. Consequently, the two qubits are jointly described by a tensor product state $\tau_{\ms{C}_0}(\beta_0) \otimes \tau_{\ms{C}_1}(\beta_1)$ and have four different energy eigenstates, i.e. $\ket{i}_{\ms{C}_0}\ket{j}_{\mathcal{C}_1}$ for $i,j \in \{0,1\}$. Let us now focus on two particular eigenstates, namely
\begin{align}
    \ket{0}_{v} := \ket{0}_{\ms{C}_0}\ket{1}_{\mathcal{C}_1}, \qquad \ket{1}_{v} := \ket{1}_{\ms{C}_0}\ket{0}_{\mathcal{C}_1}.
\end{align}
These two states have an energy spacing $\epsilon_v := \epsilon_1 - \epsilon_0$ and span a subspace of the joint Hilbert space that is usually referred to as the \emph{virtual qubit} (see also \cite{Janzing2000}). For that subspace we can further assign a \emph{virtual temperature} $\beta_v$ by looking at the ratio of populations in the virtual qubit, that is
\begin{align}
    e^{-\beta_v \epsilon_v} := \frac{{}_v\langle 1| \tau_{\ms{C}_0}(\beta_0) \otimes \tau_{\ms{C}_1}(\beta_1) |1\rangle_v}{{}_v\langle 0| \tau_{\ms{C}_0}(\beta_0) \otimes \tau_{\ms{C}_1}(\beta_1) |0\rangle_v} = \frac{e^{-\beta_0 \epsilon_0}}{e^{-\beta_1 \epsilon_1}},
\end{align}
which allows us to express $\beta_v$ as
\begin{align}
    \label{eq:virtual_temp_3qubits}
    \beta_v = \left(\frac{\epsilon_0}{\epsilon_0 - \epsilon_1}\right) \beta_0 - \left(\frac{\epsilon_1}{\epsilon_0 - \epsilon_1} \right) \beta_1.
\end{align}
Importantly, observe that the virtual temperature, as a function of the local energies $\epsilon_0$ and $\epsilon_1$, can take any range of values. In particular, notice that it can fall outside of the range specified by $\beta_0$ and $\beta_1$, and can even take negative values. This corresponds to a population inversion  \cite{Brunner2012}. 

Let us now add another qubit to the machine, denote it with $\ms{C}_z$ and place it in a thermal contact with the virtual qubit. In order to enable an interaction between the new qubit and the virtual qubit, we choose the energy of the former to be $\epsilon_{z} = \epsilon_v$. This allows the systems to resonantly exchange energy with the following Hamiltonian:
\begin{align}
    \label{eq:inter_virt}
    H_{\text{int}} =  \chi \dyad{1}{0}_{v} \ot \dyad{0}{1}_{\ms{C}_z} + \text{h.c.},
\end{align}
where $\chi$ specifies the coupling strenght and ``h.c.'' stands for Hermitian conjugate. The above interaction induces a transition between two degenerate energy eigenstates $\ket{1}_v\ket{0}_{\ms{C}_z} \leftrightarrow \ket{0}_v \ket{1}_{\ms{C}_z}$ which effectively places the virtual qubit in a thermal contact with the new qubit. After a sufficiently long amount of time the temperature of the qubit $\ms{C}_z$ reaches the virtual temperature $\beta_v$. 

Let us now observe that such a three qubit thermal machine can operate both as a refrigerator or a heat pump. In Fig. \ref{fig:regimes} we plot the virtual (inverse) temperature $\beta_v$ as a function of $\beta_1$ for a fixed value of $\beta_0$. More specifically, notice that when $\beta_1 \leq \beta_0$ the inverse virtual temperature is larger than both $\beta_0$ and $\beta_1$, hence the machine operates as a refrigerator. When $\beta_0 < \beta_1 < (\epsilon_0/\epsilon_1) \beta_0$, the inverse virtual temperature $\beta_v$ is smaller than both $\beta_0$ and $\beta_1$, hence the machine is a heat pump. Finally, when $\beta_1 > (\epsilon_0/\epsilon_1) \beta_0$ the virtual temperature is negative, meaning that the device operates as a heat engine. Notice that in all three regimes the virtual temperature falls outside of the range of ``easily accessible'' temperatures specified by $\beta_0$ and $\beta_1$ that could be achieved simply by coupling one of the qubits to the two heat baths. 
}

\subsection{Thermodynamic neuron for NOT gate} 
  \label{sec:results1}

\begin{figure*}
    \centering
    \includegraphics[width=\linewidth]{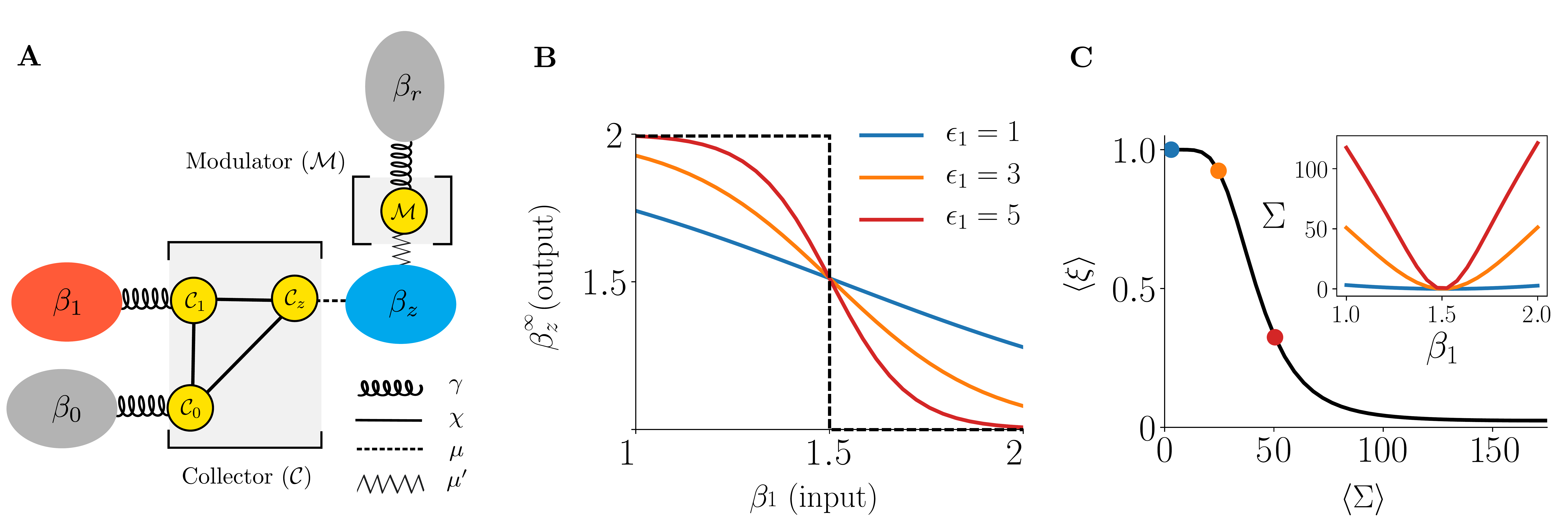}
    \caption{\textbf{Thermodynamic neuron for implementing a NOT gate.} Panel A shows the design of the machine. The collector consists of three interacting qubits (yellow dots), each connected to a thermal environment. The logical input is encoded in the temperature $\beta_1$ of the heat bath $\ms{B}_1$ (red) while the output will be retrieved from the final temperature $\beta_z^{\infty}$ of the finite-size reservoir $\ms{B}_z$ (blue); the heat bath $\ms{B}_0$ is at fixed reference temperature. The collector implements the desired inversion of the temperature. To make the response non-linear, we must add the modulator, which consists of an additional qubit connected to a reference heat bath. 
    Panel B shows the relation between the input temperature $\beta_1$ and the final output temperature $\beta_z^\infty$ (in the steady-state regime). Notably, the machine produces the desired inversion of the temperature. The quality of the response can be increased by tuning the machine parameters, in particular by increasing the energy gap $\epsilon_1$ of the collector qubit $\ms{C}_1$. \chg{Dashed black line shows the characteristics of an ideal NOT gate.} Panel C shows the trade-off between the average dissipation $\langle \Sigma \rangle$ [see Eq. (\ref{eq:avg_dissip})] and the average error $\langle \xi \rangle$ [see Eq. (\ref{eq:avg_error})]. We see clearly that in order to increase robustness to noise, the machine must dissipate more heat to the environment. The inset shows the entropy production as a function of the input temperature $\beta_1$ for different values of the qubit energy $\epsilon_1$. Parameter values: \chg{$\beta_{\text{hot}} = 1$, $\beta_{\text{cold}} = 2$}, $\gamma = \chi = 1$, $\mu = 10^{-4}$, $\epsilon_z = 0.1$, $\tau = 10^8$ and \chg{$\beta_0 = \beta_z(0) = 3/2$.}}
    \label{fig:not_gate}
\end{figure*}

In this section we describe an autonomous thermal machine implementing an inverter (NOT gate). This represents the simplest example of a thermodynamic neuron. \chg{We start with a short and intuitive description of the machine's operation after which we provide a more in-depth discussion of its functioning.} 

The machine is sketched in Fig. \ref{fig:not_gate}A. It is composed of two parts, which we refer to as the \emph{collector} $(\ms{C})$ and the \emph{modulator} $(\ms{M})$. The collector consists of three interacting qubits connected to different environments. The first two qubits (denoted $\ms{C}_0$ and $\ms{C}_1$) are connected to two heat baths denoted $\ms{B}_0$ and $\ms{B}_1$, at inverse temperatures $\beta_0$ and $\beta_1$ respectively. The first bath $\ms{B}_0$ simply represents a reference bath, hence $\beta_0$ will simply be fixed to a certain value {and called the \emph{reference temperature.}} The second bath $\ms{B}_1$ will be used to encode the input of the computation. These two heat baths are supposed to have an infinitely large heat capacity, hence their temperature will remain constant during the time evolution of the machine. Finally, the third qubit of the collector (denoted $\ms{C}_z$) is connected to an environment $\ms{B}_z$ with a finite heat capacity $C$ (this can be viewed as a finite-size reservoir). They key point is that \chg{the inverse temperature $\beta_z$ of} $\ms{B}_z$ will evolve in time, and the final temperature (in the steady-state regime) will encode the output of the computation. 

To guide intuition, it is useful to think of the collector as a simple (three-qubit) thermal machine \cite{Linden_2010,Brunner2012} \chg{which we discussed in Sec. \ref{sec:three_qubits}}.  When the input temperature is hot \chg{($\beta_1<\beta_0$)} the machine works as a refrigerator, i.e. cooling down the output environment $\ms{B}_z$. On the contrary, when the input temperature is cold \chg{($\beta_1> \beta_0$)} the machine works as a heat pump, heating up $\ms{B}_z$. Hence we see that the machine works as a sort of inverter for the temperature. \chg{We encourage the Reader to take a look at Fig. \ref{fig:regimes} which illustrates different regimes of operation of a three-qubit machine that is equivalent to the collector of the NOT gate.}

\chg{Due to the action of the collector, the output inverse temperature $\beta_z$ depends linearly on the input $\beta_1$, as demonstrated in Eq. \eqref{eq:virtual_temp_3qubits}. From a signal processing perspective, this translates to an inverting linear amplifier. When a signal passes through a sequence of such devices, any noise present in the signal will be amplified, potentially leading to unwanted bit flips. To enhance the noise robustness of the collector, a nonlinear modulation of the output inverse virtual temperature is required. This modulation should minimize the output variation for small input fluctuations within designated logical regions (i.e. where the collector acts as a refrigerator or a heat pump). In the same time, the output should change substantially when the input transitions to a different logical region. This ensures that the any noise-induced distortion of the signal in the output will be minimized.

The above modulation will be realized by another part of the machine, i.e. the modulator. It is a single qubit machine which is coupled to two thermal baths, i.e. a reference bath $\ms{B}_r$ with a fixed inverse temperature $\beta_r$ and the output bath $\ms{B}_z$ (see Fig. \ref{fig:not_gate}A).} This has the effect to delimit a specific range for the output temperatures $\beta_z$, making the response of the device effectively non-linear and hence closer to an ideal NOT gate.

\begin{figure}[t!]
    \centering
    \includegraphics[width=0.9\linewidth]{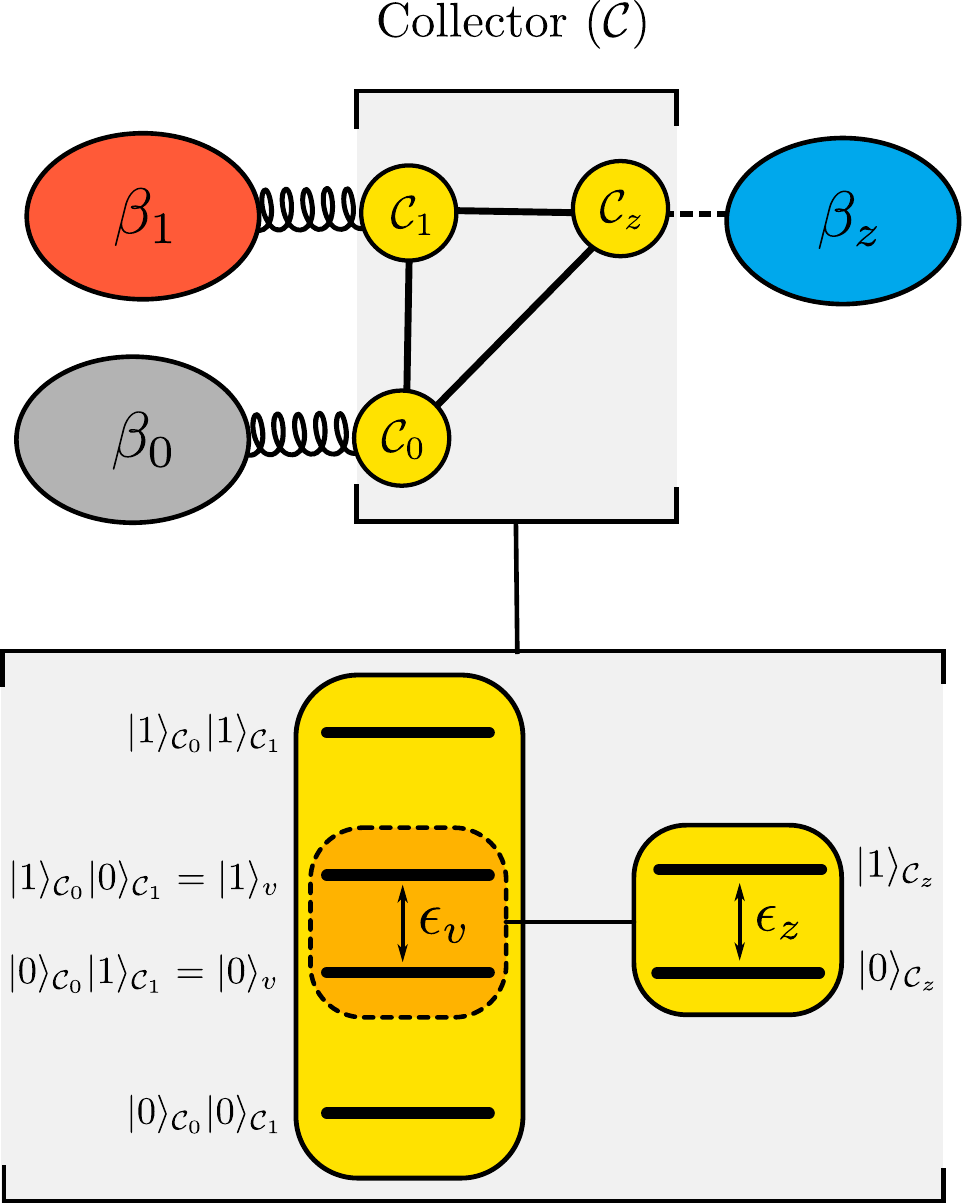}
    \caption{\textbf{Virtual qubit in the collector. } The sketch shows the energy structure of a three-qubit collector. The Hilbert space of the two physical qubits $\ms{C}_0$ and $\ms{C}_1$ contains a two-dimensional subspace with an energy gap $\epsilon_v = \epsilon_z$ (so-called virtual qubit) and effective temperature $\beta_v$ (so-called virtual temperature). The interaction Hamiltonian $H_{\text{int}}$ is chosen so that this virtual qubit interacts with the physical qubit $\ms{C}_z$, cooling it down (or heating up) in the process.  }
    \label{fig:virtual_qubit}
\end{figure}

In the following we present in detail the models for the collector and the modulator, and then discuss the dynamics of the machine and its operation as a NOT gate. Finally, we investigate the trade-off between the gate performance (as given by the average error rate) and dissipation (as given by entropy production).

\subsubsection{Collector}
The collector $\ms{C}$ is composed of three qubits which we denote $\ms{C}_i$ for $i \in \{0, 1, z\}$ (see Fig. \ref{fig:virtual_qubit}), with energy gaps $\epsilon_i$. Each qubit is weakly coupled to an environment, denoted $\ms{B}_i$, at (inverse) temperatures $\beta_i$  with the coupling constants { $\gamma$ for $\ms{C}_0$ and $\ms{C}_1$ and $ \mu$ for $\ms{C}_z$.} \chg{Therefore the collector can be seen as a three-qubit thermal machine that we discussed in Sec. \ref{sec:three_qubits}.} This three-qubit system is described by a joint state $\rho_{\ms{C}}$ that evolves according to the master equation (\ref{eq:master}), i.e.
\begin{align}
    \label{eq:lindblad}
    \dot{\rho}_{\ms{C}} = -i [H_0 + H_{\text{int}}, \rho_{\ms{C}}] + \mathcal{L}[\rho_{\ms{C}}],
\end{align}
with \chg{the local Hamiltonian} $H_0 = \sum_{i \in \{0, 1, z\}} \epsilon_i \dyad{1}_{\ms{C}_i}$
and \chg{local dissipators} \mbox{$\mathcal{L} \!=\! \mathcal{L}^{(0)} \!+\! \mathcal{L}^{(1)} \!+\! \mathcal{L}^{(z)}$}. 

It is important \chg{to ensure} that energy can flow between the qubits. For this, we choose the energy gap of the third qubit $\ms{C}_z$ to be $\epsilon_z = \epsilon_0 - \epsilon_1$. This implies that the two states $\ket{1}_{\ms{C}_0} \ket{0}_{\ms{C}_1} \ket{0}_{\ms{C}_z} $ and $\ket{0}_{\ms{C}_0} \ket{1}_{\ms{C}_1} \ket{1}_{\ms{C}_z} $ have the same energy \chg{and can be coupled via the interaction Hamiltonian}
\begin{align}
    \label{eq:2}
    H_{\text{int}} = \chi \dyad{1}{0}_{\ms{C}_0} \ot \dyad{0}{1}_{\ms{C}_1} \ot \dyad{0}{1}_{\ms{C}_z} + \text{h.c.},
\end{align}
where $\chi$ is the coupling strength. This interaction conserves the total energy (since $ [ H_0 , H_{\text{int}}]=0$), which guarantees that energy can be exchanged even in the weak coupling regime.

\chg{We want to understand the effect of the collector on the output environment $\ms{B}_z$ in the steady-state regime, i.e. when $\dot{\rho}_{\ms{C}} = 0$. To do so, we will follow the approach of Ref. \cite{Brunner2012}, which is summarized in Sec. \ref{sec:three_qubits} and visualized in Fig. \ref{fig:virtual_qubit}.}

First, note that from the form of the interaction Hamiltonian $H_{\text{int}}$, we see that there are only two states of the machine that \chg{exchange energy in} the steady-state dynamics. These are simply the two states \chg{we discussed above} that have the same energy. Now let us think of the three-qubit system as \chg{a machine comprising the first two qubits $\ms{C}_0$ and $\ms{C}_1$ and the target qubit $\ms{C}_z$. The effect of the machine is to thermalize the target qubit $\ms{C}_z$ with a virtual qubit characterized by the two levels:} 
\begin{align}
    \ket{0}_v := \ket{0}_{\ms{C}_0}\ket{1}_{\ms{C}_1}, \qquad
    \ket{1}_v := \ket{1}_{\ms{C}_0} \ket{0}_{\ms{C}_1}.    
\end{align}
These levels form a virtual qubit with energy gap $ \epsilon_v  = \epsilon_0 - \epsilon_1$. Let us denote with $g_v := \bra{1}_v\tau_{\ms{C}_0}(\beta_0) \ot \tau_{\ms{C}_1}(\beta_1) \ket{1}_v$ the occupation of the excited state of this effective system. Then, the ratio of populations in the subspace associated with the virtual qubit becomes $g_v/(1-g_v) = e^{-\beta_v (\epsilon_0 - \epsilon_1)}$, where $\beta_v$ is the (inverse) virtual temperature:
\begin{align}
    \label{eq:not_vt}
    \beta_v &= \left(\frac{\epsilon_0}{\epsilon_0 - \epsilon_1}\right) \beta_0 - \left(\frac{\epsilon_1}{\epsilon_0 - \epsilon_1} \right) \beta_1.
\end{align}
\chg{Using the intuition developed in Sec. \ref{sec:three_qubits} we can now understand the steady-state dynamics of the collector $\ms{C}$. The collector aims to thermalize the target qubit $\ms{C}_z$ to the virtual temperature $\beta_v$, as can be seen by rewriting the interaction Hamiltonian in terms of virtual qubit levels as in Eq. \eqref{eq:inter_virt}. The only difference with respect to the setting from Sec. \ref{sec:three_qubits} is that now the target qubit is itself coupled to a finite heat bath $\ms{B}_z$ at an inverse temperature $\beta_z$, and therefore the regime of the collector's operation (i.e. whether it acts as a refrigerator or a heat pump) is defined with respect to the inverse temperature $\beta_z$ instead of $\beta_0$. } 

When $\beta_v > \beta_z$ energy flows \chg{from the target qubit $\ms{C}_z$ to the machine (via the virtual qubit), effectively cooling the target qubit down;} the machine acts as a refrigerator. On the other hand, when $\beta_v < \beta_z$, energy flows towards the qubit $\ms{C}_z$, heating it up in the process; the machine acts as a heat pump. Importantly, which one of these different machine's behaviors actually occurs depends on the inverse temperatures $\beta_0$ and $\beta_1$ via Eq. (\ref{eq:not_vt}). This \chg{ability of the collector to change its behavior based on the input temperature} is the basic principle behind our inverter.

\chg{Recall that the target qubit $\ms{C}_z$ is coupled to its own (finite) thermal bath $\ms{B}_z$.} In turn, \chg{the mechanism described above will have the effect of thermalizing} the output environment $\ms{B}_z$ to the virtual temperature. To see this, consider the steady-state current from the collector $\ms{C}$ to the output environment $\ms{B}_z$ under the reset model of thermalization [see Eq. \eqref{eq:reset}]: 
\begin{align}
    \label{eq:j_not_z}
    j_{\ms{C}} := \mu \epsilon_z \left[g_z(\beta_z) - g_z(\beta_v)\right],
\end{align}
{where $g_z(x) := g(x \epsilon_z)$ and $g$ is the Fermi-Dirac distribution from Eq. \eqref{eq:fd}.} Indeed, the collector attempts to bring the temperature of the environment $\ms{B}_z$ closer to the virtual temperature. \chg{By choosing energy gaps $\epsilon_0$ and $\epsilon_1$ appropriately [i.e. the linear weights in Eq. (\ref{eq:not_vt})], we can in principle obtain any linear inverting behaviour.}

\subsubsection{Modulator}
The modulator $\ms{M}$ is composed of a single qubit with an energy gap $\epsilon_{\ms{M}}= \epsilon_z$. \chg{The qubit is put in contact with two thermal baths: $\ms{B}_r$ at an inverse temperature $\beta_r$ with a coupling rate $\gamma$, and $\ms{B}_z$ with a \chg{different} coupling rate $\mu'$.} The qubit state $\rho_{\ms{M}}$ evolves according to the following master equation:
\begin{align}
    \dot{\rho}_{\ms{M}} = \mathcal{L}^{(r)}[\rho_{\ms{M}}] + \mathcal{L}^{(z)}[\rho_{\ms{M}}].
\end{align}
In the steady-state the excited-state population of the qubit depends only on the \chg{coupling} rates $\gamma$ and $\mu'$. We set these rates so that $\mu' \ll \gamma$, ensuring that the qubit will effectively thermalize to the inverse temperature $\beta_r$. Therefore, the steady-state heat current from $\ms{M}$ to $\ms{B}_z$ \chg{under the reset model \eqref{eq:reset} reads}
\begin{align}
    \label{eq:j_not_r}
    j_{\ms{M}} := \mu' \epsilon_z \left[g_z(\beta_z) - g_z(\beta_r)\right].
\end{align}
The modulator attempts to bring $\beta_z$ closer to the (inverse) temperature $\beta_r$ and the strength of this effects is controlled by the coupling rate $\mu'$. The choice of the values of $\beta_r$ and $\mu'$ will \chg{therefore completely} specify the behavior of the modulator. By appropriately choosing these two parameters, we can \chg{specify} the range of the output temperature $\beta_z$ \chg{leading to a non-linear response of the machine (see Supplementary Material A for more details)}. 

\subsubsection{Dynamics of the machine}

We now combine our understanding of the collector and the modulator to gain insight into the full evolution of the machine. The collector and the modulator are both connected to an environment $\ms{B}_z$ with a finite heat capacity $C$. The temperature change of this environment is proportional to the sum of all entering heat currents. Specifically, we assume that \chg{the temperature $T_z := 1/\beta_z$} changes according to the calorimetric equation \chg{$\dot{T}_z = \frac{1}{C} (j_{\ms{C}} + j_{\ms{M}})$, which in terms of $\beta_z$ reads}
\begin{align}
    \label{eq:beta_dot}
    \chg{\dot{\beta}_z = -\frac{1}{C} \beta_z^2 ( j_{\ms{C}}+ j_{\ms{M}}).}
\end{align}
\chg{Consequently, the steady-state inverse temperature $\beta_z^{\infty}$ is obtained by solving the equation $j_{\ms{C}} + j_{\ms{M}} = 0$.}

Crucially, the couplings of the collector and the modulator to $\ms{B}_z$ are set to be much weaker than their couplings to the heat baths $\ms{B}_{0}$, $\ms{B}_1$ and $\ms{B}_r$, i.e. we have that $\gamma \gg \mu, \mu'$. This implies that the dynamics of the whole machine has two intrinsic time scales. The first (fast dynamics) is associated with the internal evolution of \chg{the collector} $\ms{C}$ and \chg{the modulator} $\ms{M}$. Hence, both \chg{parts of the machine} will reach their steady states relatively quickly. This means that the qubit $\ms{C}_z$ of the collector will reach the virtual temperature $\beta_v$ [see Eq. \eqref{eq:not_vt}], while the modulator qubit will be at temperature $\beta_r$. The second (slow dynamics) is associated with the changes of the temperature of the output environment $\ms{B}_z$. This means that $\ms{B}_z$ will slowly thermalize via the contact with qubits $\ms{C}_z$ and $\ms{M}$, to an intermediate temperature between $\beta_v$ and $\beta_r$.

Let us now discuss the slow evolution more carefully. We denote by $\beta_z(t)$ the time evolution of the temperature of the output environment $\ms{B}_z$. The heat currents delivered from \chg{the collector and the modulator} alter $\beta_z(t)$ according to Eq. (\ref{eq:beta_dot}). The steady-state of the output environment $\ms{B}_z$ is achieved when $\dot{\beta}_z(t) = 0$. Denoting the stationary value of $\beta_z(t)$ with ${\beta}_z^{\infty}$ and solving the equation $j_{\ms{C}} + j_{\ms{M}} = 0$, we obtain the following expression for the steady state temperature:
\begin{align}
    \label{eq:betat_ss_mt}
    g_z(\beta_z^{\infty}) = \Delta g_z(\beta_v) + (1-\Delta) g_z(\beta_r),
\end{align}
where $\Delta:=\mu/(\mu'+\mu)$. In order to interpret the temperature of the output reservoir $\ms{B}_z$ as a valid logical signal \chg{we need to limit the possible values of output temperature $\beta_z^{\infty}$ to a well-defined range $\beta_{\text{cold}}$ and $\beta_{\text{hot}}$, where the parameters satisfy $\beta_{\text{cold}} > \beta_{\text{hot}}$ but are otherwise arbitrary. In order to enforce this requirement, we can fix} the \chg{free} parameters of the modulator (see Supplementary Material A \chg{for details}). Choosing $\mu'$ and $\beta_r$ so that $\Delta = g_z(\beta_{\text{hot}})-g_z(\beta_{\text{cold}})$ and $g_z(\beta_r) = g_z(\beta_{\text{cold}})/(1-\Delta)$, leads to 
\begin{align}
    \label{eq:betaz_not_limits_mt}
    \beta_z^{\infty} = \frac{1}{\epsilon_z} \log\left[Q(\beta_v)^{-1} - 1\right],
\end{align}
with $Q(\beta_v) := g_z(\beta_{\text{hot}}) g_z(\beta_v) + g_z(\beta_{\text{cold}})(1-g_z(\beta_v))$ and $\beta_v$ is the virtual temperature given in Eq. (\ref{eq:not_vt}). 

At this point, we are ready to discuss the performance of our inverter. In Fig. \ref{fig:not_gate}B we plot the transfer characteristics (TC) of our machine in the steady-state regime. Specifically, we see that the behaviour between the input and the output temperatures, respectively $\beta_1$ and $\beta_z$, is indeed an inversion. For a cold (hot) input temperature, the output temperature is hot (cold). Note that in the figure, we have set \chg{$\beta_{\text{cold}}=2$ and $\beta_{\text{hot}}=1$}. More generally, from Eq. (\ref{eq:betaz_not_limits_mt}), we see that: ($i$) when \chg{$\beta_1 = \beta_{\text{hot}}$ we have $\beta_z^{\infty}  \approx  \beta_{\text{cold}}$, and ($ii$) when $\beta_1 = \beta_{\text{cold}}$ we get $\beta_z^{\infty} \approx \beta_{\text{hot}}$}.

Additionally, we can see from the figure that the quality of the NOT gate depends on the model parameters, in particular on the energy gap $\epsilon_1$ of the collector qubit $\ms{C}_1$. The larger $\epsilon_1$ becomes, the closer we get to an ideal NOT gate \chg{(i.e. inverted step function)}. In fact, it can be shown that, in the limit $\epsilon_1 \rightarrow \infty$, the TC becomes the ideal inverted step function. We investigate analytically in Supplementary Material A the properties of the TC in Eq. (\ref{eq:betaz_not_limits_mt}), showing its dependence on the energies of the collector qubits $\epsilon_0$, $\epsilon_1$ and the inverse temperature $\beta_0$ of the reference bath. More specifically, Eq. (\ref{eq:betaz_not_limits_mt}) describes a function which is very similar to a sigmoid (or Fermi-Dirac) function $f(x) = (1 + e^{x})^{-1}$, i.e.
\begin{align}
    \label{eq:not_response}
    \beta_z^{\infty} = f(x) +  \mathcal{O}(\epsilon_z),
\end{align}
where $x := (\epsilon_1 + \epsilon_z)(\beta_0 - \beta_1)$. When $\epsilon_z$ is small (compared to $\epsilon_1$), the roles of the free parameters become clear: $\beta_0$ characterizes the location of the step in $\beta_z^{\infty}$ and $\epsilon_0 \approx \epsilon_1$ describes its steepness. For larger values of $\epsilon_z$ the TC still demonstrates the desired inverting behavior, however the role of the parameters $\epsilon_0$ and $\beta_0$ becomes a bit more complicated to interpret (see Appendix Supplementary Material A for details).

\chg{We note that the exact functional dependence between the input $\beta_1$ and the output $\beta_z^{\infty}$ depends on the amount of heat current, and hence also on the explicit thermalization model used. To arrive at Eq. \eqref{eq:not_response} we used the simple reset model from Eq. \eqref{eq:reset}. Choosing a different  thermalization model leads to different mathematical forms (nonlinear functions $f$), however the machine's fundamental ability to invert temperatures remains unchanged.}

\subsubsection{Logic operation}
As seen above, our device produces the desired inversion relation between the input and output temperatures. The next step is to use the machine as a NOT gate, for which we must now encode the logical information appropriately in the corresponding temperatures.

In what follows the input and output signals will be described by random variables $x, y \in \{0, 1, \varnothing\}$, where $0,1$ represent the binary logical values and $\varnothing$ denotes an invalid result that cannot be assigned. The logical input $x$ is encoded in the inverse temperature $\beta_1$ of heat bath $\ms{B}_1$, while the logical output $y$ is decoded from the final (inverse) temperature $\beta_z^{\infty}$ of $\ms{B}_z$. For that we use the mapping 
\begin{align}
    \label{eq:ideal_inv_beh}
    x = \begin{cases}
    0, \hspace{5pt} \beta_1 = \beta_{\text{hot}}, \\
    1, \hspace{5pt}\beta_1 =  \beta_{\text{cold}}, 
    \end{cases} \!\!
    y = \begin{cases}
    0, \hspace{5pt} \beta_z^{\infty} \leq (1\!+\!\delta)\beta_{\text{hot}}  , \\
    1, \hspace{5pt} \beta_z^{\infty} \geq  (1\!-\!\delta) \beta_{\text{cold}},\!\! \\
    \varnothing \quad \!\text{otherwise}.\!\!
    \end{cases}\!\!
\end{align}
Parameters $\beta_{\text{cold}}$ and $\beta_{\text{hot}}$ characterize the machine's range of operation, while $\delta$ captures its robustness to noise in the output signal. \chg{All these parameters are a part of the machine's design and can be chosen arbitrarily, depending on the specific working conditions (e.g. how much noise is the machine expected to tolerate)}. Mapping logical values to intervals as above allows one to tolerate fluctuations in the output signal, i.e. interpret them correctly even if they differ between rounds due to the stochasticity of the machine's evolution. In principle we could also consider having noise in the input signal. \chg{Similarly, we could also consider mapping the output of the machine to several logic states, therefore effectively simulating a function with several output values. However, to keep the presentation simple, we will not do this here.}

\chg{The thermal machine discussed in Sec. \ref{sec:results1} performs computation in an inherently stochastic way,} and therefore the actual machine's output will fluctuate around the steady-state value from Eq. (\ref{eq:betaz_not_limits_mt}). This will lead to possible errors in the gate implementation. Characterizing these errors is important to assess the quality of the gate, in terms of its robustness to noise. 

In the following, we describe the machine as a binary channel defined by the encoding $e(\beta_1|x)$ and decoding $d(y|\beta_z^{\infty})$ as specified in Eq. (\ref{eq:ideal_inv_beh}). The input distribution is denoted $p(x)$. The behaviour of the machine is then specified by a conditional distribution 
\begin{align}
    p(y|x) := \int d(y|\beta_z) T(\beta_z|\beta_1)  e(\beta_1|x) \, \mathrm{d} \beta_1 \mathrm{d} \beta_z,
\end{align}
where $T(\beta_z|\beta_1)$ describes the actual response $\beta_z$ of the machine to the input $\beta_1$. \chg{Since the evolution is ultimately stochastic, we assume that the actual response of the machine to the input $\beta_1$ is distributed according to 
\begin{align}
    T(\beta_z|\beta_1) \propto \mathcal{N}(\beta_z^{\infty}, C),
\end{align}
where $\mathcal{N}(\mu,\sigma)$ is a Gaussian with mean $\mu$ and standard deviation $\sigma$. The output heat bath $\mathcal{B}_z$ is a macroscopic system which is composed of a large number of particles. In such a large system, according to the Central Limit Theorem, the sum of temperature fluctuations tends towards a Gaussian distribution. Since the temperature $\beta_z$ is a macroscopic property related to the average kinetic energy of the particles, it reflects the sum of these microscopic fluctuations, and hence Gaussian distribution provides a reasonable approximation to the actual distribution. Moreover, a larger heat bath (higher $C$) can sustain more energy fluctuations without a substantial change in its average temperature. This translates to a wider distribution (larger standard deviation) in the Gaussian distribution.}

The average computation error $\langle \xi  \rangle$ is the probability of observing an output different from the desired one, i.e.
\begin{align} \label{eq:avg_error}
\langle \xi \rangle &= \sum_{x \in \{0, 1\}} \sum_{y \in \{0, 1\}} p(x) p(y|x) \chg{\delta_{xy}},
\end{align}
\chg{where $\delta_{xy}$ is the Kronecker delta function.} The above quantity is directly related to the shape of the transfer \chg{characteristics} (TC), see Fig. \ref{fig:not_gate}B. Notably, the closer TC is to an ideal NOT gate (black dashed line), the smaller is $\langle \xi  \rangle$. Interestingly, the actual TC of our machine approaches the ideal one in the limit of $\epsilon_1 \rightarrow \infty$. This indicates that the quality of the computation can be enhanced at the cost of using more energy, which implies that the machine will dissipate more heat. In the following discussion, we will examine this trade-off in more detail.

\subsubsection{Trade-off between entropy production and noise robustness}

Here we investigate the relation between the quality of the gate, as quantified by the average computation error, to its thermodynamic cost, given by the amount of entropy that is produced during the computation. 

First, let us evaluate the entropy production. As mentioned the dynamics of the machine features two different time scales. The primary source of dissipation is given by the slow dynamics, in which the temperature of the output reservoir changes. The latter being connected to the collector and the modulator, the total dissipation rate is given by $\dot{\Sigma} = \dot{\Sigma}_{\ms{C}} + \dot{\Sigma}_{\ms{M}}$. We have that $\dot{\Sigma}_{\ms{C}} = -\beta_0 j_{0} - \beta_1 j_{1} - \beta_z j_{\ms{C}}$ and $\dot{\Sigma}_{\ms{M}} = -\beta_z j_{\ms{M}} - \beta_r j_{r}$; here $j_{0,1,r}$ denotes the current from the heat bath $\ms{B}_{0,1,r}$ to their respective qubit. \chg{Under the action of the} slow dynamics the entropy of the qubits in the machine does not change, i.e. $\dot{S}(\rho_{\ms{S}})= 0$. Because of this, the entropy production is the weighted sum of the heat dissipated in each environment. In order to quantify the total dissipation incurred during the computation, we have to integrate the dissipation rate over time, i.e. 
\begin{align}
\langle \Sigma(\beta_1) \rangle = \int_{0}^{\tau} \langle \Dot{\Sigma}(\beta_1) \rangle \text{d}t.
\end{align}
where $\tau$ is the running time of the computation, indicating when the final temperature output $\beta_z^\infty$ is read off. 

We see that this quantity depends on $\beta_1$. Hence the dissipation will vary depending on the input. In the inset of Fig. \ref{fig:not_gate}C, we show this beheviour, also considering different values of the parameter $\epsilon_1$. As expected, since the rate of dissipation is proportional to the heat currents flowing into the environments, the larger the energy of the qubits, the larger the rate of heat dissipation. Moreover, as expected, when $\beta_1 = \beta_0$, dissipation vanishes.

Next, let us estimate the dissipation averaged over different rounds of the computation, i.e. averaging over the inputs. We get the quantity
\begin{align} \label{eq:avg_dissip}
\langle \Sigma \rangle &= \sum_{x \in \{0, 1\}}\int\! p(x) e(\beta_1|x) \Sigma(\beta_1)\, \mathrm{d} \beta_1 .
\end{align}
In Fig. \ref{fig:not_gate}C we examine the relation between 
the total dissipation $\langle \Sigma \rangle $ and the average computation error $\langle \xi \rangle$. We consider a uniform input distribution $p(x) = 1/2$, and a sufficiently long computing time to ensure we are close to the steady-state regime, $\tau = 10^8$. There is a monotonous relation between the two quantities. As expected, we see that lowering the average error rate comes at the price of increasing the dissipation. 

\chg{Finally, let us mention that the choice of temperatures $\beta_{\text{cold}}$ and $\beta_{\text{hot}}$ plays a crucial role in the machine's performance. A wider temperature range enhances noise resistance, making the machine less susceptible to small temperature fluctuations. However, this benefit comes at a cost – a larger temperature gap increases the thermalization time, essentially slowing down the thermodynamic neuron's computations. This creates an interesting trade-off between noise robustness and computational speed.  }

\subsection{Thermodynamic neuron for linearly-separable functions} \label{sec:results2}

In the previous section we presented an autonomous thermal machine for performing a simple computation task, namely inverting a signal. In this section, we generalize this construction for performing more complex computations. In particular, we show that any linearly-separable function (from $n$ bits to one bit) can be implemented via such a machine, and give an effective algorithm for setting the appropriate machine parameters. This represents the general form of a thermodynamic neuron. 
We discuss explicitly examples for implementing the NOR gate and 3-MAJORITY.

A key step will be to establish a close connection between the thermodynamic neuron and the perceptron, the standard algorithm for modeling an artifical neuron. In particular this connection exploits the notion of the virtual qubit.

\subsubsection{Model}

\begin{figure*}
         \centering
         \includegraphics[width=\linewidth]{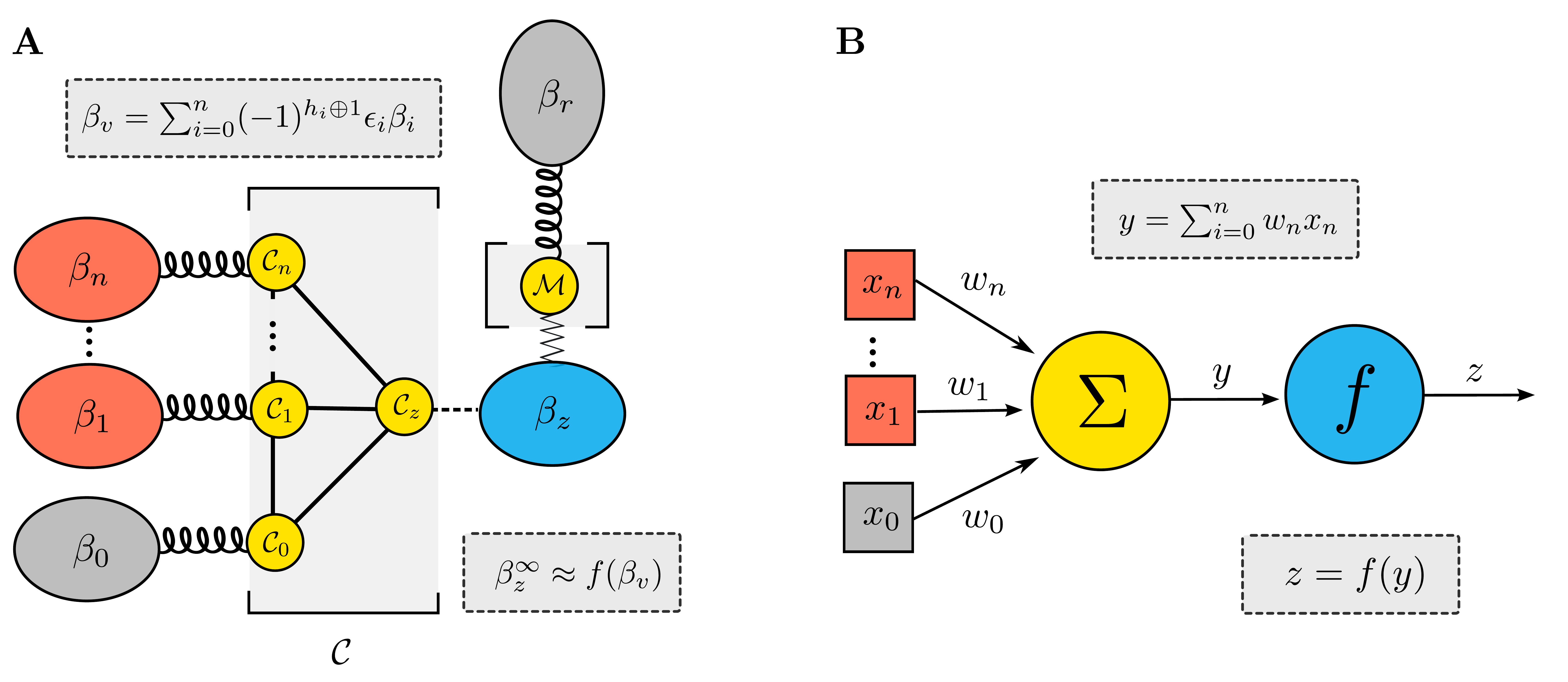}
        \caption{\textbf{General model of the thermodynamic neuron and analogy with a perceptron.} Panel A shows the structure of a thermodynamic neuron for implementing an $n$-to-one bit function. The collector $\ms{C}$ consists of $n+2$ qubits, connected to the input heat baths (red), reference heat baths (grey), as well as the output reservoir (blue). The working principle of the collector is to thermalize qubit $\ms{C}_z$ to the virtual temperature $\beta_v$ (see Eq. (29)). 
        In turn, this affects the temperature of the finite-size output reservoir $\ms{B}_z$ (blue). The modulator controls the range of output temperatures, making the response effectively non-linear. In the steady-state regime, the final output temperature is given by $\beta_z^{\infty}$ given by a non-linear function of $\beta_v$ [see Eqs \eqref{eq:beta_z_neuron} and \eqref{eq:tn_expan}]. 
        The machine can implement any linearly-separable binary function by appropriately setting the parameters: the qubit energies, the interaction Hamiltonian and the temperatures of the reference heat baths. Notably, this machine is closely connected to the perceptron model shown in panel B, which is used extensively in machine learning. Given inputs $x_k$, the perceptron first computes a weighted sum $y$, then processed via a non-linear activation (sigmoid) function $f$. Similarly, the thermodynamic neuron first creates a virtual qubit at temperature $\beta_v$, which is a weighted sum of the input temperatures $\beta_k$. Second, the modulator implements the non-linear activation function.  
        Note that in a specific regime ($\epsilon_z$ sufficiently small), the thermodynamic neuron implements a perceptron, as the activation function tends to a sigmoid in this case. 
        }
      \label{fig:3}
\end{figure*}

\chg{ A thermodynamic neuron is an autonomous quantum thermal machine that implements a binary function from $n$ bits to one bit. In analogy with the thermal machine  for inversion from Sec. \ref{sec:results1},} the general model of a thermodynamic neuron consists of two main parts, the collector $\ms{C}$ and the modulator $\ms{M}$ (see Fig. \ref{fig:3}). The design of the collector is a generalisation of the \chg{single-input collector,} while the modulator is exactly the same. 

\chg{The (generalized) collector $\ms{C}$ consists of} $n+2$ qubits $\ms{C}_i$ with energy gaps $\epsilon_i$. The first qubit $\ms{C}_0$ is connected to the \chg{reference} heat bath $\ms{B}_0$ at a fixed inverse temperature $\beta_0$. The \chg{remaining} qubits $\ms{C}_1$ to $\ms{C}_n$ are connected to input heat baths, their temperatures ($\beta_1$ to $\beta_n$) encoding the $n$ input bits. The last qubit $\ms{C}_z$ is connected to the output reservoir $\ms{B}_z$ with a finite heat capacity $C$. The modulator $\ms{M}$ consists of a single qubit connected to a heat bath at a reference temperature $\beta_r$ \chg{and} the output reservoir $\ms{B}_z$. 

\chg{In order to understand the dynamics of the collector we will again use the idea of a virtual qubit, now associated with a two-dimensional subspace within the Hilbert space of qubits $\ms{C}_0, ..., \ms{C}_n$. A multi-qubit machine can have many virtual qubits, hence we need notation to specify which virtual qubit is relevant for our problem. For that we introduce a binary vector $\bm{h} = (h_0, h_1, \ldots, h_n)$, where $h_i \in \{0,1\}$ denotes whether a given physical qubit $i$ contributes its ground $(h_i = 0)$ or excited $(h_i = 1)$ state to the virtual qubit.} 

\chg{A virtual qubit specified by a vector $\bm{h}$} consists of two \chg{multi-qubit} energy levels 
\begin{align}
    \label{eq:tp_0v}
    \ket{0}_v &:= \ket{h_0}_{\ms{C}_0} \ket{h_1}_{\ms{C}_1} \ldots \ket{h_n}_{\ms{C}_n}, \\
    \label{eq:tp_1v}
     \ket{1}_v &:= \ket{h_0\oplus 1}_{\ms{C}_0} \ket{h_1\oplus 1}_{\ms{C}_1} \ldots \ket{h_n\oplus 1}_{\ms{C}_n},
\end{align}
where $ \oplus $ denotes addition mod $2$.  The energy gap $\epsilon_v$ of the virtual qubit \chg{with levels $\ket{0}_v$ and $\ket{1}_v$} is given by
\begin{align}
    \label{eq:ev}
    \epsilon_v := \sum_{i=0}^n (-1)^{h_i \oplus 1} \epsilon_i
\end{align}
The design of the machine is then completely characterized by the vector $\bm{h}$ and the energy gaps $\epsilon_i$ for $i \in \{0, 1, \ldots, n\}$. \chg{These parameters can be chosen freely --- they specify the binary function implemented by the thermodynamic neuron.}

\chg{Let us now discuss the dynamics of the thermodynamic neuron. The machine engineers a virtual qubit at the desired temperature, and places it in thermal contact with the output qubit $\ms{C}_z$ in resonance with the virtual qubit $(\epsilon_z = \epsilon_v)$. This thermal contact is realized via an interaction Hamiltonian $H_{\text{int}} := g (\dyad{0}{1}_v \ot \dyad{1}{0}_{\ms{C}_z} + \text{h.c.}) $. In turn, qubit $\ms{C}_z$ thermalizes the output reservoir $\ms{B}_z$ to the virtual temperature.} 

\chg{To characterize the virtual temperature observe that the excited-state population of the virtual qubit in the steady-state reads}
\begin{align}
    g_v(\beta_v) := g_0(h_0) \cdot g_1(h_1) \cdot \ldots \cdot g_n(h_n),
\end{align}
where $g_i(0) = (1+e^{-\beta_i \epsilon_i})^{-1}$ and $g_i(1) = 1 - g_i(0)$. The virtual temperature $\beta_v$ satisfies $\exp[-\beta_v  \epsilon_v] = g_v(\beta_v)/(1-g_v(\beta_v))$, and is given by (see Supplementary Material B)
\begin{align}
    \label{eq:vt_perceptron}
    \beta_v = \frac{1}{\epsilon_z} \sum_{i = 0}^n (-1)^{h_i} \beta_i \epsilon_i.
\end{align}
The virtual temperature is a linear combination of input temperatures $\beta_i$, with relative weights specified by energy gaps $\epsilon_i$ and $h_i$. {This relation will be crucial in the next subsection where we establish a connection with perceptrons.}

\chg{Thermodynamic neuron, in analogy with the inverting thermal machine,} features two natural time-scales: \chg{Thermalization within the collector and the modulator happens quickly, while the thermalization of the output environment $\ms{B}_z$ happens slowly.} In particular, the time evolution of the output $\beta_z(t)$ is governed by the slow dynamics, and given by Eq.~(\ref{eq:beta_dot}).

\chg{In order to solve for the inverse  steady-state temperature $\beta_z^{\infty}$ we proceed as before [see Eq. (\ref{eq:betat_ss_mt})]. We find}
\begin{align}
    \label{eq:beta_z_neuron}
    \beta_z^{\infty} = \frac{1}{\epsilon_z} \log\left[Q(\beta_v)^{-1} - 1\right],
\end{align}
with $Q(\beta_v) := g_z(\beta_{\text{hot}}) g_z(\beta_v) + g_z(\beta_{\text{cold}})(1-g_z(\beta_v))$ and $\beta_v$ is given by Eq. (\ref{eq:vt_perceptron}). \chg{Recall} that temperatures $\beta_{\text{cold}}$ and $\beta_{\text{hot}}$ \chg{specify} the desired temperature range for the computation. Following the derivation in the previous section, we can expand $\beta_z^{\infty}$ in the energy gap $\epsilon_z$ and obtain
\begin{align}\label{eq:tn_expan}
    \beta_z^{\infty} = f(\beta_v) + O(\epsilon_z),
\end{align}
where $f(x) = (1 + e^{x})^{-1}$. Therefore, we see that for small $\epsilon_z$, the output temperature $\beta_z^{\infty}$ behaves essentially as the sigmoid function. For larger values of $\epsilon_z$ the function differs from the sigmoid one, but still offers a similar qualitative behavior. 

\chg{It is important to emphasize that the more inputs a thermodynamic neuron has, the lower is the probability of occupying its virtual subspace.  This means that the time it takes to equilibrate the target qubit $\mathcal{C}_z$ to the virtual temperature $\beta_v$ increases with the number of inputs. To address this challenge, using multiple, interconnected thermodynamic neurons arranged in a network might be more efficient than using a single, complex neuron with many inputs. We will explore how to build such networks of thermodynamic neurons in Section \ref{sec:network}.}

\subsubsection{Connection with perceptrons}

At this point, it is insightful to establish a formal connection between our model of the thermodynamic neuron and the perceptron \cite{mcculloch1943logical}. The latter represents the most common model of an artificial neuron, and serves as a fundamental component of artificial neural networks. 

The perceptron (see Fig. \ref{fig:3}B) is a simple algorithm for linear binary classification \cite{rosenblatt1957perceptron}. For a vector of inputs $\bm{x} = (x_0, \ldots, x_n)$ it produces an output $z$ given by
\begin{align}\label{eq:perc}
    z = f(y) \qquad \text{with} \quad y = \sum_{i=0}^n  x_i w_i,
\end{align}
where $x_0 = 1$ by convention, $\bm{w} = (w_0, \ldots, w_n)$ is a vector of weights that specifies the behavior of the perceptron, and $f$ is the activation function (sigmoid). The perceptron allows for a classification of the input space into two classes; it provides a linear separation of the the inputs depending on the value of the function (0 or 1).

At this point, the connection appears clearly. The thermodynamic neuron computes via a two-step procedure, that is very similar to the perceptron. First, given the inputs (encoded here in the temperatures \chg{$\beta_1$, \ldots, $\beta_n$}), the collector produces a virtual qubit, whose virtual temperature is given by a weighted sum of the input temperatures, with weights given by the energies $\epsilon_k$, see Eq. (\ref{eq:vt_perceptron}). This corresponds exactly to the computation of the weighted sum $y$ in the perceptron. Second, through the effect of the modulator, the output response becomes non-linear, and the final temperature $\beta_z^{\infty}$ is given by a nonlinear function of the virtual temperature, see Eq. (\ref{eq:beta_z_neuron}). In particular, in the regime of small $\epsilon_z$, this nonlinear functions becomes the sigmoid, hence corresponding exactly to the case of the perceptron, see Eq. (\ref{eq:tn_expan}). This analogy is important, and is further illustrated on Fig.~\ref{fig:3}.

An interesting insight from this analogy is that it sheds light on the importance of the modulator in our model. Indeed, if the machine would involve only the collector, the final output temperature would be simply the virtual temperature, corresponding to a trivial activation function $f(y) = y$ in the perceptron algorithm, which is known to perform poorly in machine learning. The modulator provides the essential ingredient of non-linearity: Its effect is to map the virtual temperature in a non-linear way \chg{to a temperature inside} the range from $\beta_{\text{hot}}$ to $\beta_{\text{cold}}$. Depending on the value of $\epsilon_z$ \chg{and the choice of the thermalization model}, we get different types of non-linear function. In particular, when $\epsilon_z$ is small \chg{and thermalization is a reset model,} we get the sigmoid function as in a perceptron. This suggests that thermodynamic neurons could serve as a physical model for a fully analogue implementation of perceptrons.

\subsubsection{Algorithm for designing the machine}

Beyond the conceptual interest, the above connection between the perceptron and our thermodynamic neuron is useful. Suppose we want to design a thermodynamic neuron implementing a given logic operation (e.g., the majority). For this one would need to find an appropriate combination of qubit energies $\{\epsilon_i\}$ and \chg{the vector $\bm{h}$ which specifies} the interaction Hamiltonian $H_{\text{int}}$. This problem is generally hard and would require rather intensive optimization, especially for more complex functions. \chg{Indeed, finding the appropriate set of parameters is equivalent to answering the following question: How to choose the systems' local and interacting Hamiltonian so that we achieve the desired steady state?} In what follows we will present a neural-network inspired algorithm which \chg{answers this question} quickly and efficiently by finding both the appropriate energy structure and the interaction Hamiltonian of the thermodynamic neuron. Notably, this structure needs to be set only once and \chg{from now on the thermodynamic neuron will serve its purpose (i.e. implement the desired function) without any further need of changing its parameters. The algorithm thus provides a general method for designing a thermodynamic neuron implementing arbitrary linearly separable functions.}

\chg{The main idea of the algorithm is to first run a classical machine-learning algorithm that finds the separating hyperplane for the (linearly-separable) binary function that one would like to implement. Then, exploiting the formal connection between the perceptron and thermodynamic neuron, one chooses the parameters of the model so that the virtual temperature directly corresponds to the separating hyperplane found by the machine-learning algorithm. }

Specifically, suppose we want to implement an $n$-input binary function $R(\bm{x})$, where $\bm{x} = (x_1, \ldots, x_n)$. First, we define the mapping between logical inputs and outputs and temperatures. The logical inputs and output are denoted with $x_1, \ldots x_n, y \in \{0, 1\}$, and encoded in the inverse temperatures of the respective environments through the following procedure:
\begin{align}
    \label{eq:ideal_inv_behII}
    x_i = \begin{cases}
    0, \hspace{5pt} \beta_i = \beta_{\text{hot}}, \\
    1, \hspace{5pt}\beta_i =  \beta_{\text{cold}}, 
    \end{cases}
    \hspace{-9pt}
    y = \begin{cases}
    0, \hspace{5pt} \beta_z^{\infty} \leq (1\!+\!\delta)\beta_{\text{hot}}  , \\
    1, \hspace{5pt} \beta_z^{\infty} \geq  (1\!-\!\delta) \beta_{\text{cold}},\!\! \\
    \varnothing \quad \!\text{otherwise}.\!\!
    \end{cases}
\end{align}
where $i \in \{1, \ldots, n\}$. As before, we focus on the range of temperatures from $\beta_{\text{hot}}$ to $\beta_{\text{cold}}$.

Next we construct a thermodynamic neuron implementing $R(\bm{x})$. For this, we must appropriately set the parameters of the machine, namely $\beta_0$, $\epsilon_k$ for $k \in \{0, 1, \ldots, n\}$ \chg{and the vector $\bm{h}$. Moreover, we also introduce a parameter $\alpha > 0$ which quantifies the overall energy scale of the qubits comprising the machine, and hence as well the quality of implementing the desired function.} For that we can use the following algorithm.

\vspace{10pt}
 \begin{algorithm}[H]
    \label{alg1}
    \DontPrintSemicolon
    \KwInput{$n$, $R(\bm{x})$, $\epsilon_z$, $\alpha$}
    \KwOutput{$\beta_0$, $\epsilon_k$ and $h_k$ for $k \in \{0, 1, \ldots, n\}$ \chg{[see  \eqref{eq:ev}]}}
\vspace{5pt}
Proceed according to the following steps:

\begin{enumerate}
    \item Construct a training set $D := \{(\bm{x}^{(i)}, y_i)\}_{i = 1}^{2^n}$, where $\bm{x}^{(i)} = (x_1^{(i)}, \ldots, x_n^{(i)})$ and $ y_i \!=\! R(\bm{x}^{(i)}).$
    \item Train a linear classifier (e.g. a sigmoid perceptron) to classify $\bm{x}_i$ into two classes: $y_i = 0$ and $y_i = 1$. This gives a vector of weights $\bm{w} = (w_0, \ldots, w_n)$.
    \item Set the elements of the vector $\bm{h} = (h_0, \ldots, h_n)$ as
    \begin{align}
        h_k = \begin{cases}
            0 \quad \text{if}\quad w_k \geq 0, \\
            1 \quad \text{if}\quad w_k < 0.
        \end{cases}
    \end{align}
    \item Set qubit energies $\epsilon_k$ as
    \begin{align}
    \epsilon_k = \begin{cases}
        \alpha|\epsilon_z - \sum_{k=1}^n w_k | \quad &\text{if} \quad k = 0,\\
        \alpha |w_k| &\text{otherwise}.
    \end{cases}
    \end{align}
    \item Set the bias inverse temperature $\beta_0$ as
    \begin{align}
        \beta_0 = \frac{|w_0|}{|\epsilon_z - \sum_{k=1}^n w_k|}.
    \end{align}
\end{enumerate}
\caption{Designing the thermodynamic neuron}
\end{algorithm}
\vspace{10pt}

To see why the above algorithm works, let us observe that the virtual temperature from Eq. (\ref{eq:vt_perceptron}) becomes
\begin{align}
    \beta_v  &= \frac{1}{\epsilon_z}\left[(-1)^{i_0}\beta_0 \epsilon_0 + \sum_{k=1}^n (-1)^{i_k} \beta_k \epsilon_k \right] \\
    &= \frac{\alpha}{\epsilon_z} \left[w_0  + \sum_{k=1}^n  w_k\beta_k \right].
\end{align}
Using the expansion from Eq. (\ref{eq:tn_expan}) we have
\begin{align} \label{eq:betaz_alg}
    \beta_z^{\infty} = f(x) + \mathcal{O}(\epsilon_z), \quad x = \alpha\left(w_0 + \sum_{i=1}^n w_k \beta_k \right),
\end{align}
which is exactly the output of the perceptron algorithm for a sigmoid activation function.  \chg{This demonstrates that the thermodynamic neuron model can implement all functions that can be realized using a (sigmoid) perceptron, namely all linearly-separable functions. In fact, the class of functions that can be implemented with a thermodynamic neuron is \emph{strictly larger} than the sigmoid perceptron, which can be seen by choosing different thermalization models.} 

{Eq. (\ref{eq:betaz_alg}) also reveals the role of parameter $\alpha$ which quantifies the steepness of the threshold separating the two outputs, or in other words, the quality of implementing the desired function. In general, $\alpha$ acts as a rescaling of all the energies $\epsilon_k$ of the qubits in the collector. Hence, increasing $\alpha$ leads to more dissipation and also lowers the errors in the computation. In particular, for the NOT gate, one can see that $\alpha = \epsilon_1$. }

\chg{Finally, we note that Algorithm \ref{alg1} should be thought of as a meta-algorithm, since it relies on a separate routine to train a linear classifier (Step 2). Consequently, its effectiveness and convergence depend on the chosen classifier's properties. Notably, using a classifier with guaranteed convergence translates to similar guarantees for Algorithm \ref{alg1}. }

To illustrate how to use Algorithm \ref{alg1} \chg{to design thermodynamic neurons} we now provide two examples. 

\subsubsection{Example 1: NOR gate}
\begin{figure}
    \centering
     \centering
        \includegraphics[width=\linewidth]{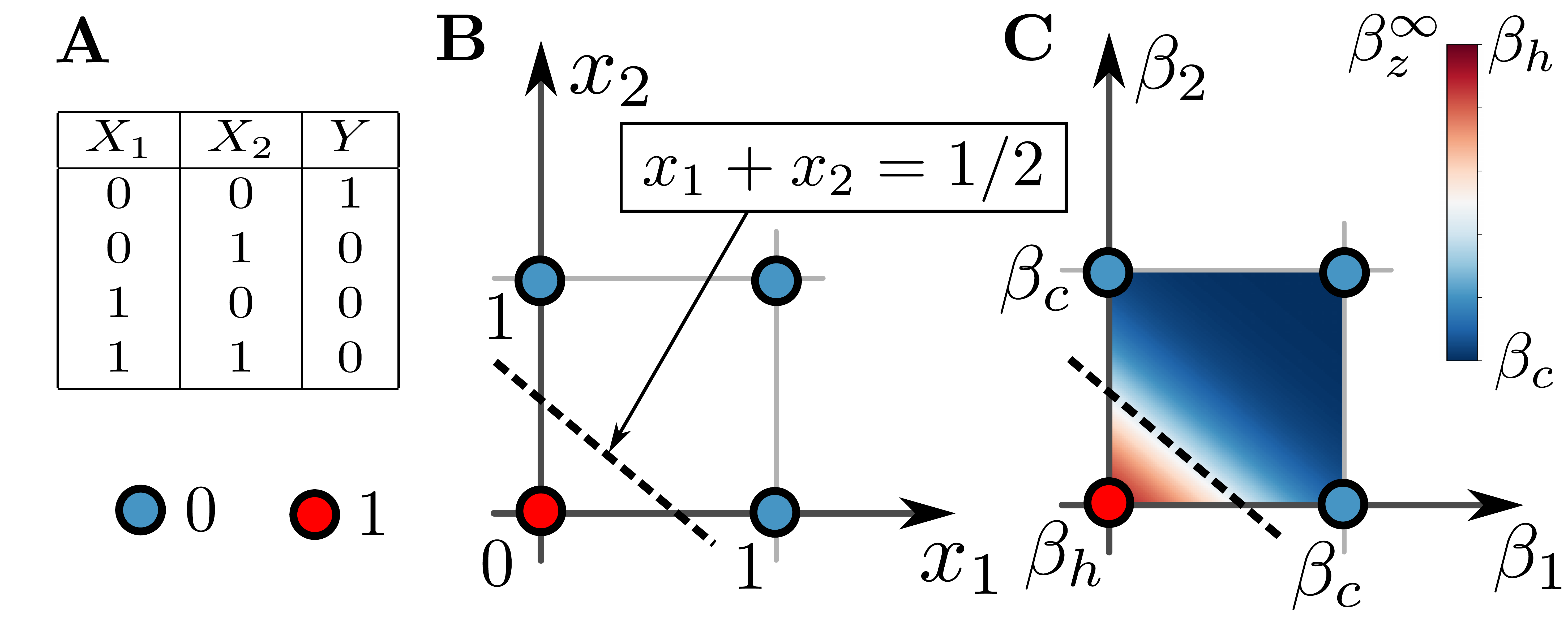}
    \caption{\textbf{Example 1: NOR.} Analysis of the thermodynamic neuron for implementing the NOR function. The truth table of NOR is given in panel A. Panel B shows all possible logical states of the machine (blue and red dots) where the colour corresponds to the desired output. Panel C shows the response $\beta_z^{\infty}$ of the thermodynamic neuron as a function of the inputs $\beta_1$ and $\beta_2$. The device does indeed implement the desired NOR gate.}
    \label{fig:5}
\end{figure}

The NOR gate takes $n=2$ input bits and returns as output the negative OR (see truth table Fig. \ref{fig:5}A). To design the thermodynamic neuron we follow the steps discussed in Algorithm \ref{alg1}. Using the truth table of NOR, we first construct the set $D$ of $2^n = 4$ data points (see Fig. \ref{fig:5}B). In principle, we could now run the algorithm and determine the vector of weights $\bm{w}$. Since in this case the separating hyperplane can be found by hand, we simply choose $x_1 + x_2 = 1/2$. This leads to the vector of weights $\bm{w} = (1, -2, -2)$. Consequently, the  interaction vector $\bm{h}$ and energy vector $\bm{\epsilon} := (\epsilon_0, \epsilon_1, \ldots, \epsilon_n)$ become
\begin{align}
  \bm{h} = (0, 1, 1), \qquad \bm{\epsilon} = \alpha (\epsilon_z+4, 2, 2),  
\end{align}
with the reference (inverse) temperature  $\beta_0 = (\epsilon_z + 4)^{-1}$.
This choice of parameters leads to the virtual temperature
\begin{align}\label{vtemp_NOR}
    \beta_v  &= \alpha (1 - 2 \beta_1 - 2\beta_2).
\end{align}
The machine's response $\beta_z^{\infty}$ is then given by Eq. (\ref{eq:vt_perceptron}) with $\beta_v$ as given above. In Fig. \ref{fig:5}C we plot the response of the thermodynamic neuron as a function of the input temperatures $\beta_1$ and $\beta_2$. The pattern of output temperatures clearly matches the desired NOR function. 

 Notably, the NOR function is functionally complete, i.e. any logic function on any number of inputs can be constructed using only NOR functions as building blocks. Consequently, by connecting multiple thermodynamic neurons appropriately one can in principle carry out any classical computation. This shows that the thermodynamic neuron is a universal.

 \subsubsection{Example 2: 3-MAJORITY}

\begin{figure}
    \centering
    \includegraphics[width=\linewidth]{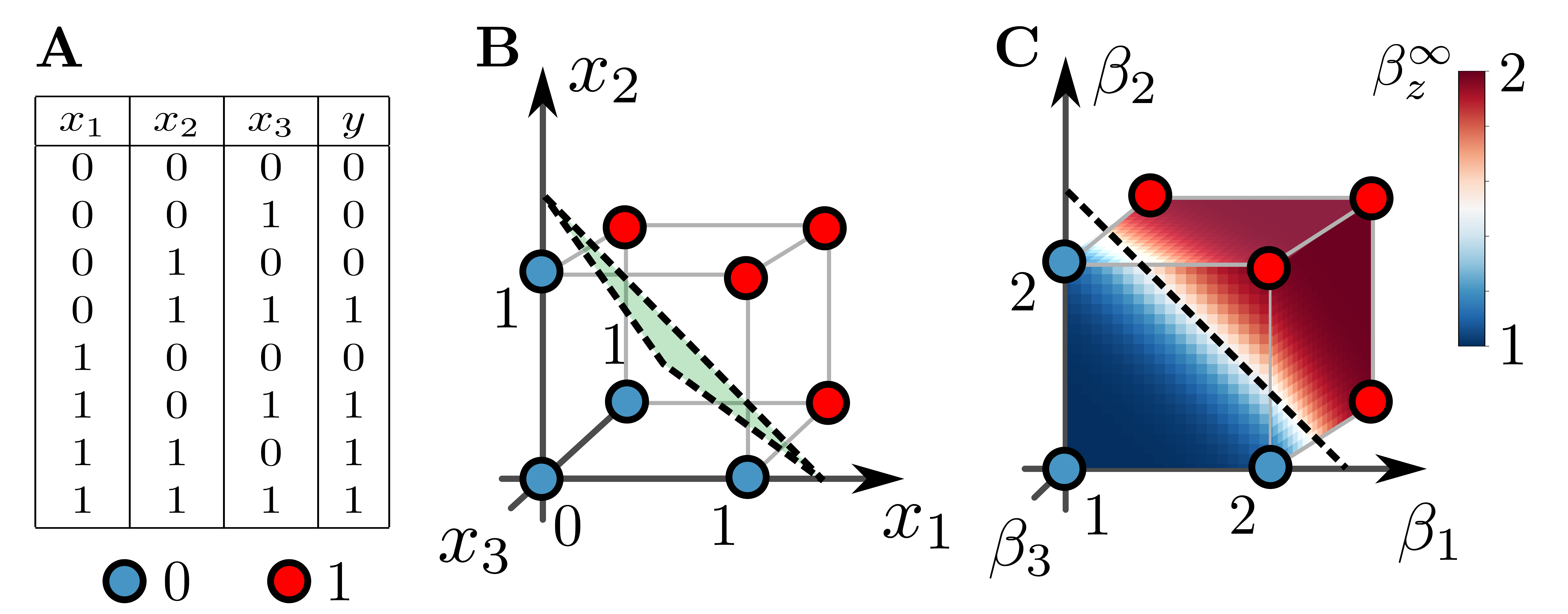}
    \caption{\textbf{Example 2: 3-MAJORITY.} Analysis of the thermodynamic neuron for implementing the majority function on three input bits. Panel A shows the  truth table. Panel B shows the possible logical states of the machine. The separating hyperplane (dashed line) is specified by the equation $x_1+x_2+x_3 = 4/3$. Panel C shows the machine's response $\beta_z^{\infty}$ as a function of the inputs $\beta_1$, $\beta_2$ and $\beta_3$. We see that the machine implements the desired operation.}
    \label{fig:7}
\end{figure}

The $3$-majority function takes $n=3$ inputs bits and outputs the majority \chg{of the input bits}. Its truth table is shown in Fig. \ref{fig:7}A. To implement 3-MAJORITY using a thermodynamic neuron we again use Algorithm \ref{alg1}. We construct the training set $D$ of $2^n = 8$ data points (see Fig. \ref{fig:7}B). Using the algorithm we found a vector of weights $\bm{w} = (-4, 3, 3, 3)$. The interaction vector $\bm{h}$ and the energy vector $\bm{\epsilon}$ are then given by
\begin{align}
  \bm{h} = (1, 0, 0, 0), \qquad \bm{\epsilon} = \alpha (\epsilon_z+12, 3, 3, 3),  
\end{align}
and the reference temperature is given by $\beta_0 = (\epsilon_z + 12)^{-1}$. This choice of parameters leads to the virtual temperature
\begin{align}
    \beta_v  &= \alpha (4 - 3 \beta_1 - 3\beta_2 - 3 \beta_3).
\end{align}
As before, the machine's response $\beta_z^{\infty}$ is given by Eq. (\ref{eq:vt_perceptron}) with $\beta_v$ specified above. In Fig. \ref{fig:7}C we plot the response of the thermodynamic neuron as a function of the input temperatures $\beta_1$, $\beta_2$ and $\beta_3$. The pattern of the output temperatures matches the desired 3-MAJORITY function.

 \subsubsection{Limitations}

From the close connection with perceptrons, we can immediately deduce a general limitation on the class of functions that can be implemented via a single thermodynamic neuron, namely linearly separable functions.

In fact, it is known that perceptron can only represent functions that are linearly separable \cite{Goodfellow-et-al-2016}. 
These are functions for which the set of inputs for which the function takes value $0$ can be separated from those whose output is $1$ via a simple hyperplane. Consequently, this constraint also limits the range of functions that can be modeled using a single thermodynamic neuron. It is however possible to overcome this limitation by considering networks of neurons. In the next section we will how networks of thermodynamic neurons can be used to compute any binary function.

\subsection{Network of thermodynamic neurons}
\label{sec:network}

Perceptrons can be assembled into a network. By increasing the complexity of such a network it gains the ability to represent more complex functions. According to the universal approximation theorem, a network with sufficiently many layers of perceptrons can approximate any binary function \cite{hornik1989multilayer}. \chg{An interesting question is whether thermodynamic neurons can also be assembled into networks in such a meaningful way. In this section we explore this question in detail.}

\chg{
\subsubsection{Combining thermodynamic neurons}
\label{sec:network_connection}

In the thermodynamic neuron the input heat baths are considered to be infinite, while the output heat bath are assumed to have a finite heat capacity. When we connect thermodynamic neurons in a network, the output of some neurons becomes the input for others. However, this poses a challenge:  How can we ensure proper functioning of the network when we treat the finite output heat bath of one thermodynamic neuron as the input to another? The finite capacity of the heat bath could disrupt the intended operation of the entire network by introducing unwanted heat currents (e.g. flowing backwards). As a result, we can no longer guarantee the validity of Eq. (\ref{eq:beta_z_neuron}) for thermodynamic neurons that constitute the network. 

A potential approach to combine thermodynamic neurons is to consider an external agent with access to infinite heat baths at temperatures $\beta_{\text{cold}}$ and $\beta_{\text{hot}}$. Let us consider a simple network composed of two concatenated thermodynamic neurons. The agent measures the temperature of the output heat bath of the first thermodynamic neuron and, depending on the outcome, couples the input qubit of the second thermodynamic neuron $(\ms{C}_1)$ to either $\beta_{\text{cold}}$ or $\beta_{\text{hot}}$. As a consequence, no unwanted heat currents flow through the output heat bath of the first thermodynamic neuron and the input qubit of the second thermodynamic neuron is coupled to an infinite heat bath. 

The proposed method for combining thermodynamic neurons relies on temperature measurements, therefore taking away their autonomy. In Supplementary Material C we present an alternative method of combining thermodynamic neurons that uses a clock. Such a device can be realized autonomously by using an autonomous clock powered by heat baths at different temperatures \cite{Erker2017}, thus providing a way to make the full computation autonomous (i.e. without invoking external control).

Based on the analysis presented above, it is evident that thermodynamic neurons can be interconnected in a manner similar to how perceptrons are linked in artificial neural networks. In this sense networks composed of thermodynamic neurons can be viewed as analog implementations of neural networks, inheriting the same capacity to perform binary functions. In other words, any function achievable by a feed-forward neural network can also be realized through a corresponding network of thermodynamic neurons. Given that neural networks are recognized for their ability to approximate any binary function, this implies that networks of thermodynamic neurons can serve as a universal model of computation.

An intriguing direction for further exploration involves considering alternative techniques for connecting thermodynamic neurons that do not necessitate extra thermodynamic resources. Moreover, one could further imagine networks of thermodynamic neurons which leverage the back-flow currents in a useful manner. This could potentially enable feedback within the network, leading to more complex and interesting network dynamics. 
}
\chg{\subsubsection{Designing networks of thermodynamic neurons}}
\chg{Finding the correct design of a network of thermodynamic neurons for implementing a given function is a non-trivial problem. In fact, there are many networks which can implement a given function. Here we discuss a heuristic approach for determining the network structure for a given binary function. We note that this is only a heuristics and hence the network of thermodynamic neurons obtained via this method is not guaranteed to implement the correct function.} 

In order to find an appropriate set of weights for a network of thermodynamic neurons we again take inspiration from artificial neural networks. More specifically, suppose we want to implement an $n$-input binary function $R(\bm{x})$. To construct the network  implementing $R(\bm{x})$ we first choose the structure of the network, i.e. the number of layers, the number of thermodynamic neurons in each layer and \chg{specify the connectivity between thermodynamic neurons}. Next we appropriately choose the free parameters of each of thermodynamic neuron, namely their reference temperature $\beta_0$, the set of energy gaps $\{\epsilon_k\}$ and the interaction Hamiltonian $H_{\text{int}}$. These parameters can be determined using \chg{a straightforward extension} of Algorithm \ref{alg1}: Basically the only difference now is that now the training step (Step $2$) is performed on the whole network rather than a single thermodynamic neuron. To illustrate this procedure, below we present a network with three thermodynamic neurons for implementing the XOR function, i.e. a function that is not linearly separable. 

 \subsubsection{Example 3: XOR gate}

\begin{figure}
    \centering
    \includegraphics[width=\linewidth]{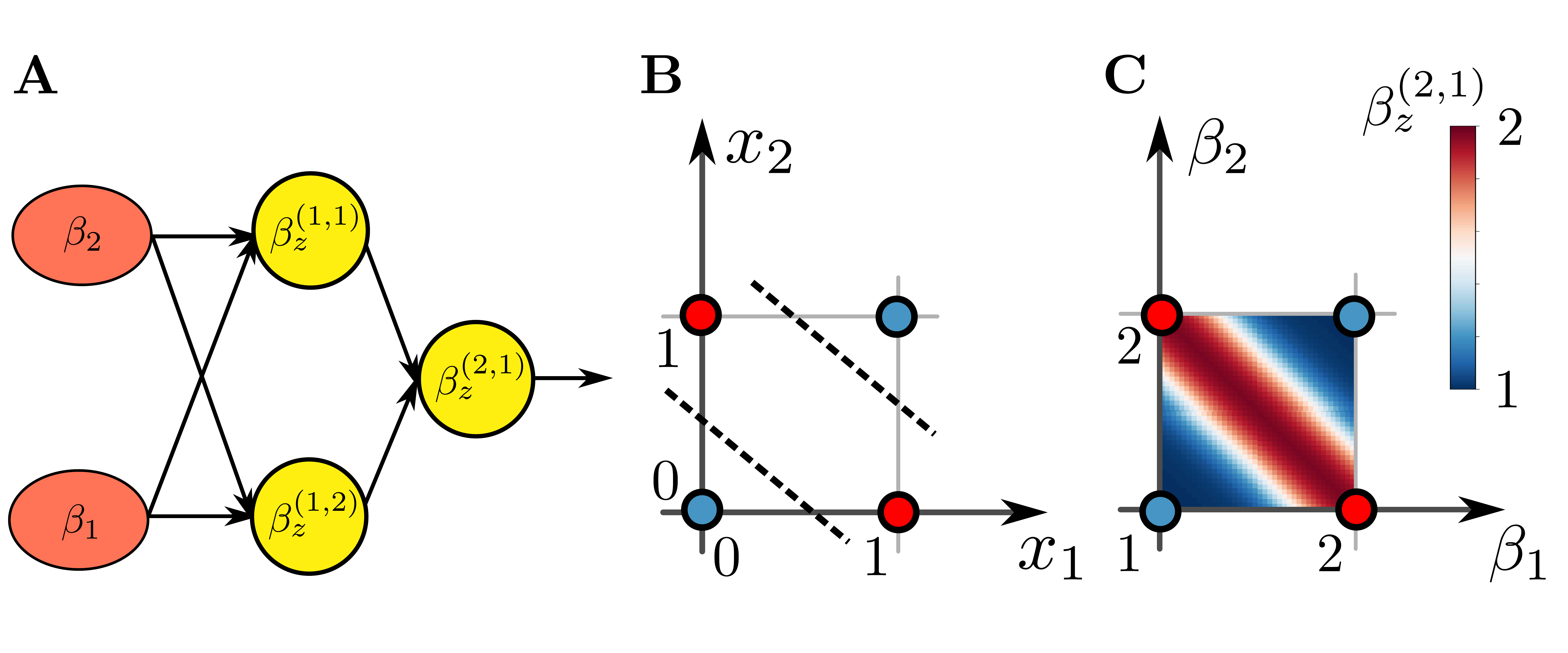}
    \caption{\textbf{Example 3: XOR.} Panel A shows the structure of a network of thermodynamic neurons that can implement the XOR function. In this case the training set \chg{(i.e. the truth table of the function for all possible inputs)} cannot be separated by a hyperplane (see Panel B), as the function is not linearly separable. The machine produces the desired response as shown in panel C: the response $\beta_z^{(2,1)}$ as a function of the inputs $\beta_1$ and $\beta_2$. Note that this machine for implementing XOR can be seen as the composition of a NAND gate and an OR gate, whose outputs $\beta_z^{(1,1)}$ and $\beta_z^{(1,2)}$ are then supplied as an input to an AND gate with output $\beta_z^{(2,1)}$. }
    \label{fig:8}
\end{figure}

The binary XOR function takes $n=2$ input bits and returns the parity. It is not a linearly separable function (see Fig. \ref{fig:8}B). Hence, it cannot be implemented with a single thermodynamic neuron. To implement XOR, we choose the network structure presented in Fig. \ref{fig:8}A. \chg{The reason for selecting this particular structure is based on the fact that a binary XOR function can be expressed as a combination of three gates, namely OR and NAND whose outputs are fed through an AND gate. The structure of the network we chose mimics this equivalence.}  Within this network structure we then use Algorithm \ref{alg1} to we compute the parameters of thermodynamic neurons implementing these three binary functions. Specifically, we construct the corresponding training set $D$ of $2^n = 4$ data points (see Fig. \ref{fig:8}B). Then we perform Step $2$ of the algorithm using the standard back-propagation algorithm \cite{rumelhart1986learning} combined with the ADAM optimization  \cite{kingma2017adam}, obtaining the vectors of weights \chg{that correspond to our approximation of the XOR function. Consequently, we use Steps $3-5$ of Algorithm \ref{alg1}} to compute the energy and the interaction vectors, as well as the reference bath temperature for each neuron. \chg{Thermodynamic neurons are then connected using the method discussed in Sec. \ref{sec:network_connection}.} The response of the machine, i.e. the inverse temperature of the last thermodynamic neuron, is shown in Fig. \ref{fig:8}C as a function of the input temperatures $\beta_1$ and $\beta_2$. We see that the network implements indeed the desired XOR function.

 \section{Discussion} \label{sec:discussion}

In this work we introduced autonomous quantum thermal machines called \emph{thermodynamic neurons} for performing classical computation. The machine is composed of several qubits which are coupled to thermal environments at different temperatures. The logical inputs and outputs of the computation are encoded in the temperatures of these environments. By engineering the energies and interactions of the machine's qubits, the device can implement any linearly-separable function. In particular, we discussed the implementation of NOT, 3-MAJORITY and NOR gates, the latter enabling universal computation. For more complex functions, we give an efficient algorithm for tuning the machine parameters. In turn, this algorithm can also be used for networks of thermodynamic neurons, which enable the direct implementation of any desired logical function. 

A notable aspect of our machines is that they rely solely on changes in temperature and energy flows: they compute with heat. \chg{This sets them apart from conventional (nano-scale) electronic computing devices and other alternative computation models, such as phonon-based computation~\cite{Ruskov2013,sklan2015splash,Lemonde2018,gustafsson2014propagating, chen2023scalable}, spintronics~\cite{wolf2006spintronics,mahmoud2020introduction,kim2022ferrimagnetic}, or superconducting circuits~\cite{pratt2023dynamical}, where heat-related effects typically hinder computation.} 

Our work also brings progress from the perspective of autonomous quantum thermal machines, by demonstrating a new application for them, namely classical computation. A single thermodynamic neuron can indeed be considered an autonomous device (see~\cite{guzmán2023divincenzolike}), while networks of them can be made autonomous via the addition of a thermodynamic clock~\cite{Erker2017}. An interesting question is whether the clock could be directly imbedded in the network of thermodynamic neurons. In parallel, our work also further demonstrates the relevance of virtual qubits and virtual temperatures \chg{for computation}~\cite{Brunner2012}. This complements recent work where these notions are used for characterizing thermodynamic properties of quantum systems~\cite{skrzypczyk2015passivity,Lipka-Bartosik2021}, the performance of thermal machines~\cite{Silva2016,Usui2021} and fundamental limits on thermodynamic processes~\cite{Clivaz2019}.

Another relevant aspect is that our model is thermodynamically consistent, in the sense of complying to the laws of thermodynamics. This allowed us to investigate its thermodynamic behavior and contrast it with the machine’s performance as a computing device. Specifically, for the NOT gate, we observe a clear trade-off between dissipation and performance, in terms of noise robustness. That is, enhancing the performance of the gate requires increasing dissipation. More generally, a similar trade-off relation between dissipation and performance exists for a general computation process carried out by the thermodynamic neuron. It would be interesting to pursue this direction further, e.g. prove a universal relationship by taking inspiration from thermodynamic uncertainty relations~\cite{seifert2019stochastic}. \chg{We want to emphasize that many models of computation consider their thermodynamic aspects under various approximations. Such approximations are generally valid in a specific range of parameters, and outside this range they can predict unphysical behavior, e.g. leading to violations of thermodynamic laws. With the growing interest in energy-efficient computing, developing thermodynamically consistent models of computation nowadays becomes increasingly important and has the potential for practical applications.}

 \subsection{Outlook} \label{sec:outlook}

Our work also opens interesting questions from the point of view of machine learning and more generally for thermodynamic computing. 

As we discussed, thermodynamic neurons have a direct connection to perceptrons and neural networks. In particular, a physical implementation of thermodynamic neurons (and more generally networks of them) would provide an alternative physics-based approach for realising neural networks. This would represent a direct (analogue) implementation, hence possibly bypassing some of the challenges of more standard digital (transistor-based) simulations of neural networks. Notably, the energy requirements and heat dissipation of the latter is very substantial, and looking for analogue implementations for reducing this thermodynamic cost is important, see e.g. \cite{Wang2023}. \chg{While the current model of thermodynamic neuron is abstract and its potential thermodynamic benefits in comparison to traditional neural network implementations are not yet well understood, investigating the relevance of the thermodynamic neurons in this context is an interesting question.}

\chg{From a more fundamental perspective, our model could also be used to investigate the thermodynamics of autonomous learning, e.g. using the techniques of Refs. \cite{keim2019memory,PhysRevX.13.031020,zhong2021machine} to modify qubit energies based on the outcome of the computation. In this way the machine would be able to ``learn'' a desired behavior in a fully autonomous manner, i.e. to improve its own decisions based on reward or penalty.} We believe this provides an interesting approach for modelling the process of learning in a thermodynamically consistent manner.

Our work can also be discussed from the perspective of thermodynamic computation \cite{Conte2023,coles2023thermodynamic,aifer2023thermodynamic}. Here, we believe that an interesting aspect of our model is the fact that computations are implemented in a physical process that is far out of equilibrium. Indeed, we use machines connected to multiple environments at different temperatures, and consider non-equilibrium steady-states. What computational power can we obtain from such a model? While we have seen that it can perform universal classical computation and is also naturally connected to neural networks, a key question is to determine its efficiency (notably in terms of time) for solving relevant classes of problems. For example, could this model provide a speed-up compared to classical computers for a relevant class of problems? 

\chg{The performance of thermodynamic neuron depends on how quickly it reaches equilibrium (thermalization). Interestingly, our simulations with a single neuron show that complete thermalization is not essential. Notably, the qualitative behavior of the model is similar even if it is allowed to thermalize only partially (so-called \emph{transient regime}). This opens exciting possibilities for exploiting the transient regime to speed up the operation of thermodynamic neurons. In the same time, full thermalization might become more important when combining multiple neurons together. On top of that, thermalization times generally increase with the number of inputs to the thermodynamic neuron. So, in some cases, using a longer network of simpler thermodynamic neurons might be a better choice than a shorter network with more complex ones. This is an interesting trade-off that we leave for further exploration in future research.}

These are rather long-term perspectives, and a more pressing one is the potential implementation of thermodynamic neurons. In this respect, recent progress on realizing autonomous quantum thermal machines with trapped ions \cite{Maslennikov2019} and superconducting qubits \cite{aamir2023thermally}, together with theoretical proposals in quantum dots \cite{Venturelli2013} and cavity QED \cite{hofer2016} are relevant. An interesting alternative is to investigate whether the physics of our model can be reproduced by a fully classical model based on rate equations. This would open the door to a classical implementation within stochastic thermodynamics \cite{Ciliberto2017}. 

\section*{Acknowldgements} We are grateful to Géraldine Haack and Paul Skrzypczyk for fruitful discussions. We also thank Nicole Yunger-Halpern, Marcus Huber, José Antonio Marin Guzman and Patrick Coles for useful comments on the first draft of this paper. We acknowledge the Swiss National Science Foundation for financial support through the Ambizione grant PZ00P2-186067 and the NCCR SwissMAP. 

\bibliography{references}

\newpage
\onecolumngrid
\appendix

\section*{Supplementary Material}

\section{Details of the thermal NOT gate}
\label{app:1}

The finite output reservoir $\ms{B}_z$ is initialised at some temperature $\beta_z(0)$ which changes according to Eq. (\ref{eq:beta_dot}). The total heat current that flows into this reservoir is a sum of two components, i.e. the currents given by Eqs. (\ref{eq:j_not_z}) and (\ref{eq:j_not_r}). More explicitly, the respective currents are given by
\begin{align}
     j_{\ms{C}} &= \mu \epsilon_z \left[g_z(\beta_z(t)) - g_z(\beta_v)\right], \\
   j_{\ms{M}} &= \mu' \epsilon_z \left[g_z(\beta_z(t)) - g_z(\beta_r)\right],
\end{align}

\begin{figure}[h!]
    \centering
    \begin{tikzpicture}[declare function={sigma(\x)=1/(1+exp((20*(\x-0.5))));
     sigma_tangent(\x)=(-1)*sigma(\x)*sigma(\x)*20*exp((20*(\x-0.5)));
     sigma_slope(\x)= 0.5 + (\x-0.5)*sigma_tangent(0.5);}]
     
\begin{axis}%
[
     axis lines=middle,
     axis line style={->},
    xlabel near ticks,
    xlabel = \(\beta_1\),
    ylabel = \(\beta_z^{\infty}\),
    grid=major,     
    xmin=0,
    xmax=1,
    ytick={0, .25, .5, .75, 1},
    ymax=1,
    samples=100,
    domain=0:1,
    legend style={at={(1,0.9)}}     
]
    \addplot[blue,mark=none]   (x,{sigma(x)});
    \addplot[domain={0.4:0.6},red,mark=none]   (x,{sigma_slope(x)});
    \addplot [dashed,mark=none] coordinates {(0.5, 0) (0.5, 1)};
     \node[] at (axis cs: 0.46,0.1) {$\beta_z^{*}$};
     \node[red] at (axis cs: 0.45,0.92) {$a$};
\end{axis}
\end{tikzpicture}
    \caption{The response $\beta_z(\beta_1)$ of the machine as a function of the input temperature $\beta_1$. The plot was generated using the following values of parameters: $\epsilon_z = 0.5$, $\epsilon_0 = 20$, $\beta_0 = 0.5$, $\beta_{\text{min}} = 0$, $\beta_{\text{max}} = 1$.}
    \label{fig:6}
\end{figure}
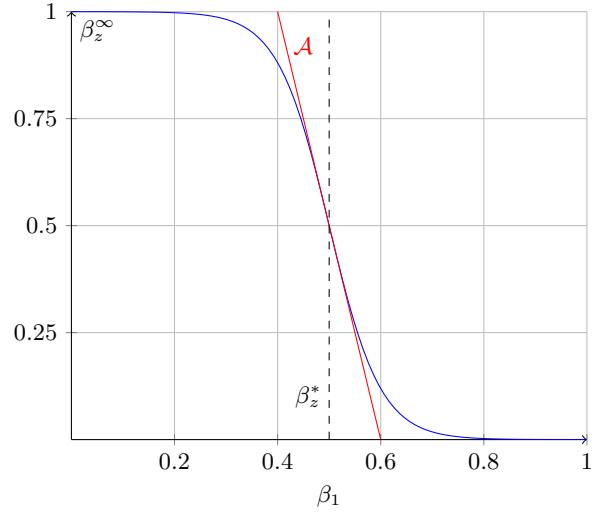

After a sufficiently long time, the finite rezervoir $\ms{B}_z$ reaches the steady-state when $\dot{\beta_z}(t) = 0$, which happens precisely when
\begin{align}
    \label{eq:betat_ss}
    g_z(\beta_z^{\infty}) = \Delta g_z(\beta_v) + (1-\Delta) g_z(\beta_r),
\end{align}
where $\Delta := \mu/(\mu + \mu')$ and $\beta_z^{\infty}$ denotes the stationary value of $\beta_z(t)$. Eq. (\ref{eq:betat_ss}) can be solved explicitly for $\beta_z^{\infty}$, i.e.
\begin{align}
    \label{eq:betaz_not}
    \beta_z^{\infty} = \frac{1}{\epsilon_z} \log\left[\frac{1}{ \Delta g_z(\beta_v) + (1-\Delta) g_z(\beta_r)} - 1\right].
\end{align}
Let us now restrict it $\beta_z^{\infty}$ to the range $[\beta_{\text{min}}, \beta_{\text{max}}]$ so that it can be interpreted as a logical signal. For that we enforce the additional constraints
\begin{align}
    \label{eq:range_constraints_not}
    \lim_{\beta_v \rightarrow \infty} g_z(\beta_z^{\infty}) = g_z(\beta_{\text{min}}), \qquad \lim_{\beta_v \rightarrow -\infty} g_z(\beta_z^{\infty}) = g_z(\beta_{\text{max}}),
\end{align}
where we recall that $\beta_z^{\infty} = \beta_z^{\infty}(\beta_v, \beta_r, \Delta, \epsilon_z)$ and $\beta_v = \beta_v(\beta_0, \beta_1, \epsilon_z)$. The additional requirements from Eq. (\ref{eq:range_constraints_not}) lead to the following set of equations.
\begin{align}
   \Delta + (1-\Delta) g_z(\beta_r) &=  g_z(\beta_{\text{min}}), \\
    (1-\Delta) g_z(\beta_r) &= g_z(\beta_{\text{max}}),
\end{align}
where $g_z(\beta_{\text{min}}) \geq g_z(\beta_{\text{max}})$. Solving these equations with substitution $\kappa := g_z(\beta_{\text{min}}) - g_z(\beta_{\text{max}}) \geq 0$ leads to $\Delta = (1- \kappa)/\kappa$ and $g_z(\beta_r) = g_z(\beta_{\text{max}}) / (1 - \kappa)$, or more precisely $\beta_r = \epsilon_z^{-1} \log [(1-\kappa) e^{\beta_{\text{max}}\epsilon} - \kappa]$. Plugging these values into Eq. (\ref{eq:betaz_not}) and solving for $\beta_z^{\infty}$ yields
\begin{align}
    \label{eq:betaz_not_limits}
    \beta_z^{\infty} = \frac{1}{\epsilon_z} \log\left[\frac{1}{g_z(\beta_{\text{max}}) + g_z(\beta_v)\kappa} - 1\right].
\end{align}
The above equation describes the steady-state response of our inverter for input $\beta_1$. We plotted $\beta_z^{\infty}$ as a function of the input temperature $\beta_1$ for the exemplary parameters in Fig. \ref{fig:6}. 

The model has three free parameters that quantify its behaviour, namely $\epsilon_0$, $\beta_0$ and $\epsilon_z$. This parameters enter Eq. (\ref{eq:betaz_not_limits}) through the virtual temperature $\beta_v$. In order to characterize the behavior of our machine we now determine some properties of the response $\beta_z^{\infty}$ as given by Eq. (\ref{eq:betaz_not_limits}). The threshold value $\beta_z^*$ can be computed by finding the root of $\partial_{\beta_1}^2 \beta_z^{\infty} = 0$, i.e. $\beta_z^* := \arg_{\beta_1} \left(\partial_{\beta_1}^2 \beta_z^{\infty} = 0\right)$, which gives 
\begin{align}
    \beta_z^* = \beta_0 \left(1 + \frac{\epsilon_z}{\epsilon_0}\right) + \frac{1}{2\epsilon_0}\log \left[\frac{1 + \cosh{\left(\beta_{\text{min}} \epsilon_z\right)}}{1 + \cosh{\left(\beta_{\text{max}} \epsilon_z\right)}}\right].
\end{align}
For small $\epsilon_z$ we have that $\beta_z^{*} \approx \beta_0$, therefore in this regime $\beta_0$ specifies the threshold temperature for which the machine changes its regime of operation. Another interesting characteristic of the function $\beta_z^{\infty}$ is the slope $\mathcal{A}$ at the threshold point, i.e.
\begin{align}
    \label{eq:slope}
    \mathcal{A} := \frac{\partial \beta_z^{\infty}}{\partial \beta_1}\Bigg|_{\beta_1 = \beta_z^*} = -\frac{\epsilon_0}{e_z} \cdot \frac{g_{\text{max}} + g_{\text{min}} - 2 g_{\text{max}} g_{\text{min}} + 2 (1-g_{\text{min}}) g_{\text{min}} \sqrt{\frac{g_{\text{max}}(1-g_{\text{max}})}{g_{\text{min}}(1-g_{\text{min}})}} }{g_{\text{max}} - g_{\text{min}}},
\end{align}
where we used a short-hand notation $g_{\text{max}} := g_z(\beta_{\text{max}})$ and $g_{\text{min}} := g_z(\beta_{\text{min}})$. For example, for a particular choice of parameters $\beta_{\text{min}} = 0$ and $\beta_{\text{max}} = 1$ we have $\mathcal{A} = -( \epsilon_0 / e_z) \tanh\left({\epsilon_z}/{4}\right) = -\epsilon_0/4 + \mathcal{O}(\epsilon_z^2)$. We therefore see that, for small $\epsilon_z$, the parameter $\epsilon_0$ specifies the slope of the threshold in $\beta_z^{\infty}$. 

\section{Details of the thermodynamic neuron model}
\label{app:2}
Consider the $n+1$ qubits $\ms{C}_i$ for $i \in \{0, 1, \ldots n\}$ that comprise the collector $\ms{C}$ with energies arranged in a vector $\bm{\epsilon} = (\epsilon_0, \epsilon_1, \ldots \epsilon_n)$ and weakly coupled to heat baths with corresponding temperatures $\beta_i$ arranged in a vector $\bm{b} = (\beta_0, \beta_1, \ldots, \beta_n)$. The logical action of the thermodynamic neuron is characterized by a string $\bm{s}$ with elements $\pm 1$ defined as $\bm{s} := [(-1)^{h_0 \oplus 1}, (-1)^{h_1\oplus 1}, \ldots, (-1)^{h_n \oplus 1}]$. The energy of qubit $\ms{C}_z$ is chosen to be $\epsilon_z = (\bm{s}-\bar{\bm{s}} )\cdot\bm{\epsilon} = \sum_{k} (-1)^{h_k} \epsilon_k$. The steady state solution for $\beta_z^{\infty}$ satisfies 
\begin{align}
    \label{eq:betat_ss_perceptron}
    g_z(\beta_z^{\infty}) = \Delta g_z(\beta_v) + (1-\Delta) g_z(\beta_r),
\end{align}
The virtual temperature $\beta_v$ satisfies
\begin{align}
    e^{-\beta_v \epsilon_z} = \frac{g_0(h_0) g_1(h_1) \ldots g_n(h_n)}{g_0(h_0\oplus 1) g_1(h_1 \oplus 1) \ldots g_n(h_n\oplus 1)} \hspace{5pt} \implies \hspace{5pt} \beta_v = \frac{1}{\epsilon_z} \sum_{i=0}^n \log \left[ \frac{g_i(h_k\oplus 1)}{g_i(h_k)} \right] = \frac{1}{\epsilon_z} \sum_{k = 0}^n (-1)^{h_k} \beta_k \epsilon_k.
\end{align}
Proceeding as in Appendix \ref{app:1} we can now restrict the range of $\beta_z$ to $[\beta_{\text{min}},\beta_{\text{max}}]$ by demanding that Eq. (\ref{eq:range_constraints_not}) is satisfied, remembering that now $\beta_v$ is a linear combination of $n+1$ temperatures. We therefore arrive at the following expression for $\beta_z^{\infty}$:
\begin{align}
    \beta_z^{\infty} = \frac{1}{\epsilon_z} \log\left[Q(\beta_v)^{-1} - 1\right],
\end{align}
where $Q(\beta_v) := g_z(\beta_{\text{hot}}) g_z(\beta_v) + g_z(\beta_{\text{cold}})(1-g_z(\beta_v))$ and $\beta_v$ is the virtual temperature given in Eq. (\ref{eq:not_vt}). 
From here one can perform similar types of calculations as in Appendix \ref{app:1}. 

\chg{\section{Combining thermodynamic neurons using a clock}
\label{app:3}
An alternative approach to combine thermodynamic neurons into networks is to use a timing device (a stopwatch clock). The main observation is that we can ensure correct operation of the device when the layers of thermodynamic neurons are synchronized, i.e. are operating one after the other. 

Let us define two relevant time parameters in this context. Firstly, we denote with $t_{\text{steady}}$ the time required for the first thermodynamic neuron to reach a temperature that is sufficiently close to its steady state output temperature $\beta_z^{\infty}$. Secondly, let $t_{\text{const}}$ represent a time interval during which the temperature of the finite thermal environment $\ms{B}_z$ can be considered constant when used as the input for the second thermodynamic neuron. These two time parameters can be determined numerically by explicitly solving the machine's dynamics. Crucially, they both depend on the chosen thermalization model.

The gateway operates by successively turning on and off the coupling between the finite heat bath $\ms{B}_z$ and the qubit $\ms{C}_1'$. In the first iteration the coupling is turned off, hence the first thermodynamic neuron reaches its steady state as if it was not connected to any other system. After time $t_{\text{steady}}$ the coupling between $\ms{B}_z$ and $\ms{C}_1'$ is turned on for time $t_{\text{const}}$ and the second thermodynamic neuron evolves towards its steady-state. In this time range the heat bath $\ms{B}_z$ effectively behaves as an infinite heat bath. After time $t_{\text{steady}} + t_{\text{const}}$ the coupling is turned off and the first thermodynamic neuron is to drive the finite bath $\ms{B}_z$ back to the target temperature $\beta_z^{\infty}$. The cycle is repeated roughly $t_{\text{therm}} / t_{\text{const}}$ times. Consequently, at each time the two thermodynamic neurons operate effectively as two disconnected devices: No unwanted heat currents flow through $\ms{B}_z$ and qubit $\ms{C}_1'$ is coupled to a heat bath which for all purposes behaves as an infinite heat bath. 

The above dynamics requires a timing device which will turn on and off the qubit-bath couplings at appropriate times. Such a dynamics can also be realized autonomously by using an autonomous clock powered by heat baths at different temperatures \cite{Erker2017}. Therefore the advantages of clock-based gateway is that it does not require measuring the temperatures of heat baths and the network can (when using an autonomous clock) operate fully autonomously, i.e. without the need of a time-dependent control. The downside of this approach is that it requires a ticking clock (stopwatch) that significantly increases the complexity of model and leads to additional thermodynamic costs. We emphasize that this implementation of a gateway allows for further optimization: By choosing different thermalization models one can vary $t_{\text{steady}}$ and $t_{\text{const}}$. 
}

\end{document}